# In vivo imaging of central nervous system fluid spaces using synchrotron radiation-based micro computed tomography

Marta Girona Alarcón[1,+], Willy Kuo[1,+], Mattia Humbel[2], Christine Tanner[2], Luca Fardin[3], Britta Bausch[1], Yann Decker[4], Irene Spera[5], Griffin Rodgers[2], Hans Deyhle[2], Alberto Bravin[6], Masato Hoshino[7], Arash Panahifar[8,9], Kentaro Uesugi[7], Sergei Gasilov[10], Petr Pleskač[11], Yuansheng Zhang[1], Diane de Zélicourt[1], Amandine Brenna[11], Ahmad Kamal Hamid[12], Pooya Razzaghi Khamesi[1], Britta Engelhardt[11], Steven T. Proulx[11], Bert Müller[2], Vartan Kurtcuoglu[1,13,14,*]

[1]   University of Zurich, Department of Physiology, The Interface Group, Zurich, Switzerland

[2]   University of Basel, Department of Biomedical Engineering, Biomaterials Science Center, Allschwil, Switzerland

[3]   University College London, Department of Medical Physics and Biomedical Engineering, London, United Kingdom

[4]   Saarland University Medical Center, Department of Neurology, Homburg, Germany

[5]   University of Basel, Department of Biomedicine, Basel, Switzerland

[6]   University of Milano-Bicocca, Department of Physics, Milano, Italy

[7]   Japan Synchrotron Radiation Research Institute, Spectroscopy and Imaging Division, Sayo, Japan

[8]   Canadian Light Source, Biomedical Imaging and Therapy Beamline, Saskatoon, Canada

[9]   University of Saskatchewan, College of Medicine, Department of Medical Imaging, Saskatoon, Canada

[10]  Helmholtz-Zentrum Hereon, Institute of Material Physics, Hamburg, Germany

[11]  University of Bern, Medical Faculty, Theodor Kocher Institute, Bern, Switzerland

[12]  Broad Institute of MIT and Harvard, Imaging Platform, Cambridge, United States

[13]  University of Zurich, Neuroscience Center Zurich, Zurich, Switzerland

[14]  University of Zurich, Zurich Center for Integrative Human Physiology, Zurich, Switzerland

[+] Equal contribution       [*] Corresponding author

## Abstract

Current approaches to *in vivo* imaging of the mouse central nervous system (CNS) do not offer a combination of micrometer resolution and a whole-brain field of view. To address this limitation, we introduce an approach based on synchrotron radiation-based hard X-ray micro computed tomography (SRµCT). We performed intravital SRµCT acquisitions of mouse CNS fluid spaces at three synchrotron radiation facilities. Imaging was conducted on both anesthetized free-breathing and ventilated animals, with and without retrospective cardiac gating. We achieved whole-brain imaging at 6.3 µm uniform voxel size, observed the distribution of cerebrospinal fluid (CSF) contrast agent over time and quantified choroid plexus movement. SRµCT bridges the gap between multiphoton microscopy and magnetic resonance imaging, offering dynamic imaging with micrometer-scale resolution and whole-organ field of view. Intravital SRµCT will play a crucial role in validating and integrating hypotheses on CSF dynamics and solute transport by providing unique data that cannot be acquired otherwise.

# Main

Central nervous system (CNS) fluids contribute to brain and spinal cord homeostasis. The interstitial fluid of the brain and spinal cord parenchyma, and the cerebrospinal fluid (CSF) within the ventricles and subarachnoid space facilitate the removal of metabolites, serve as conduits for immune cells, and act as pathways for signaling molecules[1,2]. Historically, the larger CNS fluid spaces – including the cerebral ventricles, cisterns, and the spinal subarachnoid space – have been primarily studied using physiologic techniques in animal models such as rabbits, cats, dogs, and sheep[3]. The advent of magnetic resonance imaging (MRI) greatly accelerated research on CSF dynamics in human subjects[4]. Concurrently, advances in genetic engineering shifted much of neurobiology research towards mouse models, offering powerful tools for studying CNS function and disease[5,6].

While current MRI scanners are well-suited for probing the comparably large structures of the human CNS, even the most advanced high-field, small-bore units reach their limits in terms of resolution and signal-to-noise ratio for *in vivo* imaging of the much smaller fluid spaces in mice[7–9]. The early 2000s saw the adoption of multiphoton fluorescence microscopy for studying mouse CNS fluid spaces[10–12], a technique that, despite its sub-micrometer in-plane resolution, is limited by a small field of view and shallow penetration depth, typically restricting imaging to regions close to the brain surface[13,14].

This course of technology development has led to a compartmentalization of data on CNS fluids and their interactions with surrounding tissues[15–18]. The lack of continuous data across scales contributes to the advancement of unproven models of CNS fluids, particularly when gaps are filled with assumptions rather than evidence[19]. As mouse models will continue to be central to neurobiology research in the foreseeable future, there is a critical need to bridge the gap between the localized imaging provided by multiphoton microscopy and the lower resolution whole-brain imaging offered by MRI.

To meet this need, we introduce an approach employing synchrotron radiation-based micro computed tomography (SRµCT) for *in vivo* imaging of mouse CNS fluid spaces, providing whole-brain coverage at micrometer-scale resolution. While intravital SRµCT has previously been applied in mice, it has not been used for imaging the CNS fluid spaces[20–28], most likely due to substantial technical barriers. Our method directly addresses these, enabling dynamic intravital imaging of mouse CNS fluid compartments. Synchrotron radiation has also been employed in 2D radiographic studies[21,29–37], for imaging in other animal models[38–42] and for *ex vivo* imaging of biological tissues[33,43–48]. However, these prior studies either did not involve 3D tomography, were not conducted *in vivo*, or did not image mice – and none of them targeted the CNS fluid spaces.

At the time of the experiments, only four synchrotron radiation facilities worldwide were both equipped for intravital applications and met the technical requirements for micrometer-scale, full-field imaging of mouse CNS fluid spaces. In this study, we demonstrate the effectiveness and adaptability of our approach through its successful implementation at three of these facilities. Our experimental setups and protocols have been refined for use across synchrotron radiation platforms and are openly documented to facilitate adoption by further groups. Intravital SRµCT, as presented here, can provide comprehensive insights into CSF dynamics and CSF space anatomy, with whole-brain coverage at micrometer-scale resolution, and allows real-time observation of fluid movement, fluid-tissue interactions, and physiologic changes occurring *in vivo*.

# Results

## Development of an SRμCT setup for *in vivo* imaging of mouse CNS fluid spaces

Unlike conventional computed tomography, which relies on X-ray tubes to produce the radiation required for imaging, SRμCT utilizes better collimated and more coherent X-rays generated by relativistic electrons circulating in the storage ring of a particle accelerator. The superior photon flux enables the use of monochromators to select a narrow energy band, which can be tuned to the absorption edge of a contrast agent. It also permits much faster acquisitions of radiographs, with exposure times in the 5–15 ms range – well below the duration of a mouse's cardiac and respiratory cycles. This capability is crucial for gated imaging and allows for complete 3D tomographic scans within 10–32 s, in contrast to the hours required by conventional laboratory X-ray sources to achieve comparable signal-to-noise ratios.

At the time of this study, four beamlines worldwide met the requirements for *in vivo* imaging of mouse CNS fluid spaces: the Biomedical Beamline ID17 at the European Synchrotron Radiation Facility (ESRF, Grenoble, France), the BL20B2 beamline at the Japanese Super Photon ring - 8 GeV (SPring-8, Sayo, Japan), the Biomedical Imaging and Therapy (BMIT) beamline at the Canadian Light Source (CLS, Saskatoon, Canada), and the Imaging and Medical Beamline (IMBL) at the Australian Synchrotron (AS, Melbourne, Australia). We initially developed the core methodology at ESRF ID17, subsequently refining and testing it at SPring-8 BL20B2 and CLS BMIT. Planned experiments at AS IMBL were aborted due to flooding during our allocated beamtime.

Since beamlines must support a wide range of experimental applications, users are often required to bring their own equipment. This necessitates the use of portable, rapidly deployable setups to make best use of the limited access time (beam time) allocated for experiments. Our *in vivo* SRμCT setup is designed with this constraint in mind and is built around a radio-transparent, heated mouse holder with multi-point fixation (Fig. 1). Each of these features is critical: radio-opaque materials in the beam path and unintended animal motion can produce imaging artifacts, while inadequate thermal regulation may lead to hypothermia, jeopardizing physiologic stability and potentially forcing early termination of the experiment. To promote reproducibility and facilitate broader adoption of *in vivo* SRμCT, we have made all necessary scripts, hardware designs, and procedure templates available in a Zenodo repository[49].

In contrast to intravital multiphoton microscopy, where the experimenter and the imaged animal are in proximity, SRμCT necessitates spatial separation between the imaging hutch and the control room to ensure radiation safety. Therefore, monitoring of vital signs, reactions to changes in those, and the initiation of experimental protocols must be carried out remotely. Our setup allows remote monitoring of body core temperature, blood oxygenation levels, end-tidal $CO_2$ partial pressure, and cardiac activity by electrocardiography (ECG). It also enables remote control of inhalation and injection anesthetics, infusion of contrast agent, adjustment of mouse holder temperature, and synchronization of image acquisition with ventilation and/or cardiac action.

Flexibility is a key requirement for *in vivo* SRμCT setups, as each synchrotron radiation facility and beamline present a unique set of constraints. Substantial variations arise from site-specific control systems for beam operations, which are especially relevant when synchronizing image acquisition with the animal's cardiorespiratory signals. Other technical differences, such as beamline layouts that affect cable routing or variations in mechanical interfaces to tomography rotation stages, are more easily addressed but equally important. Further challenges stem from differences in local legislations, such as accepted electrotechnical standards or permitted anesthetics, all of which may affect the comparability of results across facilities.

## Anatomic imaging at cellular resolution and whole-organ field of view

To demonstrate the utility of SRµCT for intravital imaging of mouse CSF spaces, we acquired both native-state and contrast-enhanced tomographic scans of the entire mouse brain at 6.3 µm and 8.0 µm uniform voxel size. For contrast-enhanced imaging, we infused a barium-based nanoparticle contrast agent into the right lateral cerebral ventricle. This markedly increased the contrast between CSF and surrounding tissues. In one region of interest within the dorsal third ventricle, the contrast-to-noise ratio (CNR) rose from 0.2 before infusion to 7.9 after infusion. The enhanced contrast enabled precise delineation of the ventricles, choroid plexus, and the cranial and spinal subarachnoid spaces (Fig. 2a, b). It also allowed for the tracing of CSF outflow pathways, such as those extending along the olfactory bulbs (Fig. 2b, left).

X-ray contrast agents typically incorporate elements with strong attenuation properties, attributed to their high atomic numbers and electron densities – such as iodine, barium, gadolinium, or gold. Various formulations exist based on these elements, and the optimal choice depends on the specific application. Prior to selecting the contrast agent used in this study, we screened multiple commercially available candidates based on their technical specifications, narrowing the selection to six candidates suitable for intravital use (Supplementary Information). The candidates were administered to mice *in vivo* and then evaluated through post-mortem imaging at ESRF ID17, using photon energies above their respective K-edges. The very narrow X-ray energy bandwidth achievable using monochromators at synchrotron radiation facilities enabled us to maximize the contrast between anatomic regions with and without contrast agents. Among the contrast agents tested, two colloidal nanoparticle suspensions provided the highest CNR. Of these, ExiTron™ nano 12000 exhibited lower aggregation tendencies than Aurovist™ 15 nm, and was therefore selected for use in the *in vivo* experiments.

Native-state imaging also allowed for differentiation between CSF and surrounding tissue, albeit with reduced contrast (Fig. 2c). To improve CNR between materials with similar attenuation properties, we applied Paganin's propagation-based phase retrieval filter[50]. The resulting denoised, phase-retrieved images exhibited improved area contrast, but also showed pronounced streak artifacts caused by strongly attenuating materials or sharp density gradients at the boundaries of highly X-ray absorbing structures, such as bone (Fig. 2c, right). To evaluate the filter's effect on the boundary between CSF and tissue (Fig. 2d), we compared image intensity profiles with and without phase-retrieval along a line crossing the ventricle-parenchyma interface (Fig. 2e). The improved CNR in the phase-retrieved image facilitated clearer distinction of fluid–solid boundaries, though at the cost of reduced image sharpness. This trade-off is due to the low-pass filtering nature of the Paganin algorithm. The filter strength can be modulated by adjusting the $\delta/\beta$ ratio, allowing dataset-specific optimization of the balance between CNR and spatial resolution. For this dataset, we used $\delta/\beta = 200$. When tuned in this manner, however, the $\delta/\beta$ ratio no longer reflects intrinsic material properties and loses physical interpretability[51].

For quantitative imaging, SRµCT absorption contrast images acquired with monochromatic X-rays can be reconstructed such that the image intensity values correspond directly to linear attenuation coefficients. This requires knowledge of the photon energy and effective voxel size, as well as an experimental setup and imaging strategy that minimize artifacts not related to absorption, such as streaking, local tomography artifacts, and phase effects. Subsequently, as attenuation scales linearly with contrast agent concentration, the linear attenuation coefficients can be mapped to concentration values using a calibration series of reference solutions (Fig. 2f, g). For instance, in the yellow-bordered region between the olfactory bulbs (Fig. 2b), the mean linear attenuation coefficient measured $(0.6 \pm 0.1)$ cm$^{-1}$ at 40 min after infusion start, corresponding to a mean Ba concentration of $(21 \pm 4)$ mg/ml. Mean barium concentrations in the right lateral ventricle (green-bordered region), infusion cannula (blue), fourth ventricle (cyan), and spinal subarachnoid space (SAS, magenta) were $57 \pm 4$, $191 \pm 13$, $52 \pm 4$, and $26 \pm 4$ mg/ml, respectively.

The most challenging aspect of *in vivo* SRµCT is carrying out procedures on live animals in environments where animal experimentation is not routinely conducted, and within the strict time constraints imposed by the competitively allocated beamtime. The existence of *ex vivo* imaging modalities – such as histology or tissue clearing combined with selective plane illumination microscopy – that offer concurrently high resolution and large fields of view may prompt the question of whether *in vivo* acquisitions are strictly necessary for anatomic studies. To address this, we sought to illustrate the extent of morphologic changes that occur in internal CSF fluid spaces peri-mortem. We acquired a time series of native-state tomographic images of the ventricular system in a mouse euthanized with an overdose of anesthetics, acquiring one tomographic scan every 40 s for 40 min (Supplementary Video 1). Cessation of vital functions occurred 19 min after scan start, preceded by motion artifacts resulting from deep, unconscious terminal breaths. Progressive ventricular shrinkage was first evident around 10 min later, i.e. 29 min after scan initiation.

For a quantitative analysis of ventricular shrinkage, we imaged the brain of an animal before and after euthanasia by pentobarbital injection, following prior *in vivo* infusion of contrast agent. A reduction in ventricular size was observed immediately after death (4 min after injection). Semi-automatic segmentation revealed an overall ventricular shrinkage of about 37% in the total segmented volume, with the largest absolute reduction occurring in the left lateral ventricle (from 3.0 mm$^3$ to 1.9 mm$^3$), and partial ventricular collapse upon death (Fig. 3a). Contraction occurred in all spatial directions, albeit to varying degrees. Prior to euthanasia, both lateral ventricles had volumes of 3.0 mm$^3$. Post-mortem, the right lateral ventricle, into which contrast agent had been infused, showed a relative shrinkage of 24%, whereas the contralateral (left) ventricle shrank by 38%. The third ventricle exhibited a 45% volume reduction (from 1.7 mm$^3$ to 1.0 mm$^3$). Boundaries between the lateral ventricles, third ventricle, cerebral aqueduct, and fourth ventricle were defined based on visible anatomic landmarks in the imaging data, guided by reference to the Allen mouse brain atlas[52].

To quantify changes in the cerebral aqueduct – a conduit with a disproportionate contribution to flow resistance in the ventricular spaces – we isolated the aqueduct within a smaller region of interest (Fig. 3b) and performed a dedicated segmentation to compare fine-scale changes in selected coronal cross-sections (Fig. 3c). The largest differences in cross-sectional area ($A$) and hydraulic diameter ($D_h$) between the live and post-mortem states were observed near the ventricular boundaries (Fig. 3d). At the rostral junction with the third ventricle, $A$ decreased from 0.018 mm$^2$ (live) to 0.005 mm$^2$, while $D_h$ decreased from 129 to 64 µm. At the caudal junction with the fourth ventricle, $A$ decreased from 0.059 to 0.030 mm$^2$, while $D_h$ decreased from 118 to 71 µm. These changes became more pronounced in the directions extending further into the third and fourth ventricles (Fig. 3e, f). In contrast, the narrow central portion of the aqueduct exhibited minimal to no quantifiable change in cross-section (Fig. 3d).

## Mapping spatiotemporal solute distribution throughout the cranial CSF space

To demonstrate how SRµCT can elucidate solute transport dynamics within the CSF space, we acquired time series consisting of one tomographic scan every 30 s during and after contrast agent infusion into either the right lateral ventricle or the cisterna magna (Fig. 4a, b). To minimize disruption of physiologic CSF flow, we selected low infusion rates and volumes: 0.2 µl/min for a total of 1 µl over 5 min for intra-cerebroventricular infusion, and 0.5 µl/min for a total of 2.5 µl over 5 min for intra-cisterna magna infusions. In comparison, the CSF production rate in mice is in the range of 0.3-0.7 µl/min[53,54]. During the infusion phase, contrast agent transport is likely dominated by the action of the infusion pump. After infusion ends, however, distribution of the agent is governed by the natural interplay of diffusion and convection, driven by endogenous CSF production, ependymal ciliary motion, and potentially cardiorespiratory action and muscle activation[55,56].

Contrast agent concentration at the infusion site in the right lateral ventricle followed a four-phase temporal pattern. During infusion, the concentration initially remained low, followed by a quick ramp-up beginning approximately 3 min after infusion start (Fig. 4c, panel 2, curve 1, cyan). In the second phase, concentration rose rapidly, reaching a peak at approximately 9 min. This was followed by a decline over roughly 9 min at approximately half the rate of the initial rise. In the final phase, beginning at 18 min after infusion start, concentration decreased more gradually and at a relatively steady rate. These dynamics were reflected in the corresponding linear attenuation coefficient values, which showed a peak rise rate of 1.18 cm$^{-1}$min$^{-1}$, and a prolonged decline at approximately -0.04 cm$^{-1}$min$^{-1}$ during the final phase. A similar temporal pattern was observed in this phase at sampling locations in the left lateral ventricle, contralateral to the infused right lateral ventricle (Fig. 4c, panel 2, curve 4, green). The caudoventral segment of the right lateral ventricle displayed slower concentration dynamics than the infusion site during the first two phases, but showed a more rapid decline between 18 and 33 min (Fig. 4c, panel 2, curve 2, blue). Supplementary Video 2 provides a movie of the coronal rostral plane shown in Fig. 4d, while Supplementary Video 3 contains projections of the complete time series.

During infusion into the cisterna magna (CM), we observed a rapid initial rise in contrast agent concentration at the infusion site 2 min after infusion start, reflected by a linear attenuation coefficient increase rate of 2.78 cm$^{-1}$min$^{-1}$, reaching a maximum of 4.54 cm$^{-1}$ at 3 min after the start of infusion (Fig. 4c, panel 4, curve 1, orange). This was followed by a similarly rapid decline to approximately zero within about 2.5 min (Supplementary Video 4). The temporal profile was markedly different from the more gradual concentration decrease observed in the third phase of ventricular infusion (Fig. 4c, panel 2, curve 1, blue). The use of a metal infusion-cannula, as opposed to the plastic cannula used for ventricular infusion (Fig. 4f), produced streak artifacts. These artifacts were less pronounced at the cannula's thinner tip positioned near the imaging region of interest than at its broader base, which was located farther from the imaging field (Fig. 4d, right panel).

The images shown in Fig. 4 were acquired with a uniform voxel size of 6.45 µm. To estimate the effective spatial resolution, i.e., the minimum distance at which two objects can be resolved within the same image, we computed the Fourier shell correlation (FSC). Owing to the Nyquist limit, the theoretical minimum resolvable distance in digital imaging is twice the voxel length, or 12.9 µm in this case. Resolution was calculated as the inverse of the spatial frequency at which the FSC curve fell below the 1-bit threshold[57]. The estimated spatial resolution was 19.3 µm for the intra-ventricular infusion dataset and 19.2 µm for the intra-cisterna magna dataset, corresponding to 2.99 and 2.98 voxel lengths, respectively (Fig. 4e). FSC curves for the other datasets presented in this work are available in Supplementary Figure 2. The resolution of the images acquired at CLS BMIT was 22.7 µm (2.84 voxel lengths), while those acquired at SPring-8 BL20B2 were 17.5 µm (2.19 voxel lengths) and 16.5 µm (2.07 voxel lengths), approaching the Nyquist limit.

## Quantifying tissue motion

To demonstrate the utility of intravital SRµCT for capturing physiologic motion in both intracranial and extracranial tissues, we assessed periodic displacements using retrospectively cardiac-gated tomography and non-periodic displacements using time series of non-gated acquisitions. One extracranial structure that deformed in synchrony with cardiac activity was the nasopharynx, located between the tympanic bullae laterally, the basisphenoid bone superiorly, and directly above the soft palate, which is not visible in computed tomography without specific contrast enhancement (Fig. 5a, b). This motion was identified through visual inspection of images reconstructed at 10 ms intervals following the ECG R-peak (Supplementary Video 5). The cross-sectional area of the nasopharynx increased by up to 2% between 10 and 150 ms post-R-peak – the two most decorrelated phases in the cardiac cycle. For visualization, we employed image subtraction, which also confirmed that the motion was not due to bulk displacement of the head (Fig. 5c): stationary structures such as the skull faded into the white background, while the nasopharynx and hyoid bone remained visible in a blue-red divergent colormap indicating image intensity differences. The cyclic deformation of the hyoid bone is consistent with its anatomic positioning within soft tissue near the pulsating carotid arteries.

To quantify the observed tissue motion, we semi-automatically segmented the nasopharynx from reconstructions using projections acquired at 10 and 150 ms after the R-peak, and generated a surface displacement map (Fig. 5d-g). The segmented region extended from the rostral end, beginning at the transition from the nasopharyngeal meatus surrounded by the hard palate, to the caudal limit at the level of the laryngeal inlet. Overlays of the smoothed surface meshes at both time points are shown in Fig. 5d and 5e. The surface displacement was subtle: in the magnified view shown in Fig. 5f, the surface-to-surface distance at the lateral site marked with arrows measured 8 µm. To visualize motion across the entire nasopharynx surface, we computed the point-to-point surface displacement from each vertex to its nearest neighbor on the surface at the second time point (Fig. 5g). The resulting displacement map revealed asymmetric motion, with predominant displacement toward the soft palate (Fig. 5g), and minimal to no displacement on the opposing side facing the basisphenoid bone (Fig. 5f). This pattern is consistent with the difference images from reconstructed slices (Fig. 5c).

Speckle-like artifacts visible in the 3D displacement map likely reflect minor segmentation variability arising from image noise and streak artifacts in the underlying reconstructions.

In this experiment, we lacked both a ground truth for nasopharyngeal displacement and anatomic landmarks with sufficiently high positional accuracy to benchmark motion detection limits. Estimating the lower bound of detectable motion is likewise challenging, as it depends on a combination of factors, including the effective pixel size (6.3 µm in our case), the contrast of image features, and the specifics of the motion quantification pipeline. Previous studies have reported subpixel motion detection capabilities, e.g., down to approximately 10% of the pixel size in dynamic phase-contrast radiography[58,59] and around 25% in post-gated µCT[60]. However, we anticipate a higher detection threshold in this setting due to the presence of noise and artifacts, the non-rigid nature of the observed deformation, and the difficulty of establishing precise point correspondence on the relatively smooth, tubular surface of the nasopharynx.

To assess non-cyclic tissue motion, we also acquired a time series of non-gated, standard tomographies. Within these time series, movement of the choroid plexus (ChP) was observed (Fig. 5h-n and Supplementary Video 6). The ChP is a filamentous tissue located within the ventricles. Anchored to the ventricular walls and suspended in CSF, it can undergo passive displacement in response to external forces such as CSF flow. During the 50 min observation period, the lateral ventricular ChP exhibited the greatest displacement. Subtraction images in coronal and horizontal planes comparing two representative time points, spaced 1 min apart, revealed discernible shifts of the ChP both in the left and right lateral ventricles (Fig. 5h and 5j, respectively, with anatomic locations shown in Fig. 5i). When comparing tomographic acquisitions separated by 2 min, the observed displacement was even more pronounced. We quantified three-dimensional motion between the two time points using surface-based displacement maps for the left and right lateral ventricles (Fig. 5l and 5m, respectively). The displacement patterns differed between the two locations, with the left ChP exhibiting greater displacement (median: 27 µm) than the right (3 µm), as illustrated in the corresponding histogram (Fig. 5n). Note that a logarithmic scale is used in Fig. 5l–n.

## Discussion

Synchrotron radiation-based X-ray micro computed tomography (SRµCT) enables *in vivo* 3D imaging of mouse CNS fluid spaces with a combination of spatial resolution and field of view not readily attainable with other modalities. As implemented here, SRµCT provided whole-brain coverage at a uniform voxel size of 6.3–8 µm, achieving resolution approximately one order of magnitude higher than that typically used in *in vivo* MRI for small animals[9], and about one order of magnitude lower than that attainable with intravital multiphoton microscopy. The latter, however, is typically limited to superficial cortical regions and offers roughly three orders of magnitude smaller volumetric coverage[61].

Both native-state and contrast-enhanced imaging can be performed effectively, enabling visualization of anatomy, solute dynamics, and tissue motion across the entire brain. Acquisition times of 10–32 s for *in vivo* whole-brain 3D imaging position SRµCT ahead of both conventional multiphoton microscopy, where a typical *in vivo* measurement at depth would require hours for the same number of voxels covering a much smaller field of view[62], and MRI, which typically demands several minutes to hours depending on resolution[63,64]. These comparably short acquisition times for whole-organ imaging at micrometer-scale resolution make SRµCT particularly well suited for addressing questions that require anatomic continuity across the entire brain. Even shorter acquisition times down to 1 ms are technically possible with SRµCT[65], albeit at the expense of spatial resolution, field of view and signal-to-noise. By filling this methodological gap, intravital SRµCT offers a valuable complement to established approaches, especially in studies where high-resolution structural and dynamic information must be acquired simultaneously under physiologic conditions.

While this combination of capabilities distinguishes SRµCT from other imaging modalities, the technical demands of this method make it challenging to implement. Transitioning from conventional, locally available µCT scanners to synchrotron radiation sources involves leaving a controlled environment optimized for small-animal imaging and entering a general-purpose facility where climatic conditions are often suboptimal for maintaining physiologic stability. These challenges are exacerbated when animals spend extended periods under anesthesia due to preparatory surgical procedures, such as tracheotomy and cannulation of CSF spaces, or during long time-series acquisitions. In such settings, larger animals tend to fare better, as they are less susceptible to hypothermia and dehydration over time. An additional layer of complexity arises when moving from 2D radiography to 3D tomography, which necessitates rotating the animal along with all attached supply lines and cables, while ensuring stable fixation. Taken together, these barriers may explain why intravital SRµCT of the mouse CNS fluid spaces has not been pursued earlier.

To enable such *in vivo* imaging, we developed a modular setup centered on a radio-transparent, heated mouse holder with multi-point fixation. The mouse holder design is open, freely available, and optimized for brain SRµCT. Beyond providing mechanical stability and thermal control, our system integrates remote monitoring and management of vital parameters, synchronization of image acquisition with artificial ventilation and cardiac cycle, and automated contrast agent infusion. These remote capabilities are essential for maintaining physiologic stability given the spatial separation between the imaging and control areas at synchrotron radiation facilities. The setup further accommodates technical variability across beamlines, such as differences in cable routing, rotation stage geometries, and control software. Its successful deployment at three independent synchrotron radiation facilities demonstrates the adaptability and robustness of the system.

An important advantage of SRµCT over multiphoton imaging is its ability to provide visualization of CNS fluid spaces without the use of contrast agents or fluorescent tracers, and consequently without requiring surgical intervention. In our native-state acquisitions, the intrinsic contrast between soft tissue and CSF was sufficient for delineating major anatomic compartments. Such data can be used, for example, to produce detailed structural maps of the ventricular system and associated spaces under physiologic conditions. Contrast-enhanced imaging improved fluid-tissue contrast, thereby simplifying segmentation and aiding interpretation in regions susceptible to artifacts or characterized by inherently low signal-to-noise ratios. Native-state and contrast-enhanced imaging can also be combined within an experimental series, with native-state acquisitions providing a baseline unaffected by the introduction of contrast agents. Using this approach, we observed substantial morphological changes in the ventricular system peri-mortem, underscoring the importance of intravital imaging for accurate anatomic characterization and highlighting the need for careful consideration when interpreting post-mortem datasets quantitatively.

To investigate solute distribution dynamics, we performed calibrated, contrast-enhanced SRµCT imaging, enabling quantitative estimation of solute concentrations throughout the cranial CSF system over time. The resulting concentration maps can be used to gain insights into the interplay between diffusion and convection within CNS fluid spaces. While MRI can also be employed for such analyses – and typically achieves a lower minimum detectable contrast agent concentration – accurate absolute quantification is more challenging due to the non-linear relationship between concentration and signal intensity[66], and inherently lower spatial and temporal resolutions. In contrast, the linear quantification capabilities of SRµCT offer a comprehensive, organ-wide view of solute movement and provide a robust basis for data-driven and model-based flow analysis.

SRµCT further enabled us to visualize and quantify subtle tissue motions. Although not initially intended for this purpose, cardiac gating revealed periodic deformations of extracranial soft tissues such as the nasopharynx and hyoid bone, synchronized with the cardiac cycle and likely driven by arterial pulsations. These results also indirectly confirmed the proper functioning of our gating procedure. Periodic brain motion resulting from cardiovascular action is recognized as one of the main drivers of CSF movement in humans[67,68], and it is hypothesized to also play a role in the mouse[13]. Motion measurements across the entire human brain are typically performed using MRI, which inherently suits such analyses, as it does not rely on clearly visible or trackable surfaces. While dedicated acquisition sequences for brain motion imaging are available for clinical scanners[69], to our knowledge, comparable sequences do not yet exist for small-animal scanners. This leaves SRµCT as a complementary tool for investigating brain motion in mice. In our experiments, however, we did not detect cyclic displacement at the ventricular wall-CSF interface, possibly due to insufficient spatial resolution relative to the amplitude of brain motion, temporal blurring of movement by averaging across >2 000 cardiac cycles, interference from respiration-induced motion, or a true absence of cardiovascular-driven brain motion in mice[56].

In contrast, we did observe non-periodic motion of the choroid plexus (ChP) using sequential tomographic acquisitions. The ChP, a delicate filamentous structure suspended within ventricular CSF, showed measurable displacements on the order of tens of microns, with side-specific differences indicating heterogeneous mechanical coupling to ventricular fluid dynamics. Although the relatively slow ChP movements matched well with the temporal resolution achievable with consecutive tomographic acquisitions, capturing faster non-periodic motion would require alternative approaches such as contrast-enhanced radiography. Nonetheless, the capacity of SRµCT to quantify three-dimensional tissue motion at high spatial resolution holds promise for studies aimed at elucidating the mechanical environment of the brain and understanding the forces that drive CSF flow under various physiologic and pathologic conditions.

A broader limitation, in addition to the specific ones mentioned above, is that intravital SRµCT relies on synchrotron radiation facilities equipped with beamlines suitable for tomographic imaging and appropriate animal infrastructure. At present, three such beamlines with the technical specifications required for imaging mouse CNS fluid spaces are operational – in Australia, Canada, and Japan – following the discontinuation of this imaging capability at ESRF. Additional facilities in Germany and the United Kingdom are in the planning or concept phase, respectively. This scarcity typically requires users to travel from afar and bring their specialized equipment. Although our modular setup substantially reduces the associated logistical burden, such experiments remain inherently more challenging and expensive than using a dedicated multiphoton microscopy setup in one's own laboratory or a small-animal MRI or µCT scanner nearby. Furthermore, even if increased demand stimulates the commissioning of additional beamlines, experimental time allocation will remain competitive and limited. Therefore, rather than seeking to replace established technologies, the SRµCT approach presented here fills a distinct methodological gap: simultaneously providing comparatively high spatial and temporal resolution with whole-brain coverage in living mice.

In summary, we have developed an approach for intravital imaging of mouse CNS fluid spaces using SRµCT. It provides access to whole-organ volumetric data at spatial and temporal resolutions on the order of 10 µm and 10 s, respectively. We anticipate that this method will be valuable for validating and refining models of CNS fluid dynamics and solute transport, particularly in the context of testing hypothesis generated from multiphoton datasets or computational simulations.

## Methods

An overview of the experimental and imaging parameters for the presented datasets is provided in Supplementary Table 1. Individual animals are identified by their *SubjectID* in the Supplementary Information, the Zenodo repository and the FABRIC4 data portal[49,70].

### SRµCT setup for *in vivo* imaging of mouse CNS fluid spaces

In SRµCT, where the X-ray source is stationary, radiographic acquisition from multiple angles is achieved by rotating the sample around an axis perpendicular to the beam using a motorized rotation stage. *In vivo* imaging thus requires a dedicated setup to positionally stabilize the animal and provide life support and monitoring during rotation.

At the core of our setup is a modular two-piece mouse holder. The headpiece was additively manufactured using an Objet30 Pro 3D printer (Stratasys, Edina, United States) with VeroClear™ photopolymer, chosen to provide high resolution and smooth surface finish. The headpiece provides three-point fixation via two adjustable ear bars and a bite bar integrated with a nose cone. The half-cylindrical body piece was manufactured by selective laser sintering of polyamide 12 using a Formiga P 110 printer (EOS, Munich, Germany), ensuring mechanical stability. Internal conduits allow warm water circulation to maintain core body temperature, with temperature-controlled water supplied by a circulation thermostat placed outside the imaging hutch (Haake 000-3350, Thermo Fisher Scientific Inc., Karlsruhe, Germany).

The mouse holder was mounted to the rotation stage via an aluminum adapter plate of 200 mm (ID17 and BMIT) or 300 mm (BL20B2) diameter, depending on the available physical space. The plate's larger diameter relative to both the holder itself and the rotation stage base plate provided additional space to mount a PhysioSuite® capnograph (Kent Scientific, Torrington, United States) near the animal's exhalation tubing, while preventing cabling from getting caught in moving components. Cables, ventilation tubing, and warm water lines were routed through a drag chain to prevent drooping, intertwining or encroachment into the field-of-view during rotations of up to 360°.

Artificial ventilation was provided by a SAR-1000 rodent ventilator (CWE Inc., Ardmore, United States). Separate inspiration and expiration lines were connected to a Y-piece tracheal cannula positioned near animal's lung to minimize dead space. The expiration line was split to allow parallel pressure monitoring via the SAR-1000 and capnography via the PhysioSuite device. The PhysioSuite system also continuously recorded core body temperature via a rectal probe, as well as oxygen saturation and heart rate via a pulse oximeter.

The ventilator was supplied with either oxygen or oxygen-enriched air. Safety parameters were selected based on established guidelines from the literature[71,72]. To generate a positive end-expiratory pressure of 5 $cmH_2O$, expiratory gases were passed through a water column before release. The ventilator operated in pressure mode, i.e., inspiration proceeded until either the peak inspiratory pressure limit (15-16 $cmH_2O$) or the inspiration time limit (35% of the respiratory cycle) was reached. The respiratory rate was set to 100-120 breaths per minute, and the inspiratory flow rate to 125 ml/min.

Ventilation trigger signals and pressure waveforms were read from the ventilator's analogue BNC output and digitized via a DT9808 data acquisition interface (Measurement Computing Corp., Norton, United States) connected to a computer via USB. Data were recorded using custom software implemented in LabVIEW (Version 21.0.1f2, National Instruments, Austin, United States) or LabChart (ADInstruments, Dunedin, New Zealand). The same setup was used to record a three-lead electrocardiogram (ECG) using a Bio Amp FE23 amplifier (ADInstruments, Dunedin, New Zealand), with electrode inserted into the two front paws and the tail. Where applicable, camera acquisition trigger signals from the beamline were also recorded. Capturing all signals through the same acquisition hardware and software minimized temporal synchronization issues.

Cardiac and respiratory cycles could optionally be synchronized by triggering the ventilator based on the ECG signal. For this setup, the analogue output of the electrocardiograph analogue output was routed via a BNC splitter to an Arduino Mega 2560 microcontroller (Arduino, Monza, Italy) running custom software. The Arduino generated analogue trigger signals, which were sent to the ventilator's DE-9 D-subminiature input port.

Infusions of contrast agent or injection anesthetics during imaging were carried out using KDS Legato 130 syringe pumps (KD Scientific Inc, Holliston, United States). The syringe pumps, PhysioSuite, and data acquisition interface were connected via 10 m active USB 2.0 extension cables, routed into the imaging hutch through labyrinth cable ducts integrated into the radiation shielding. This configuration enabled remote control and real-time monitoring during imaging. Syringe pumps were operated using Pump Terminal software (Version 1.0.5, Harvard Bioscience Inc, Holliston, United States), and PhysioSuite data were read out via Free Serial Port Terminal software (Version 1.0.0.710, HHD Software Ltd., London, United Kingdom).

## Anatomic imaging at micrometer-scale resolution and whole-organ field of view

### Selection of contrast agent

All *in vivo* experiments involving contrast agent utilized ExiTron[TM] nano 12000 (Viscover[TM], nanoPET Pharma GmbH, Berlin, Germany), a preclinical X-ray contrast agent composed of barium-based nanoparticles with a nominal diameter of 110 nm[73–75]. This agent was selected following a comparative evaluation of six commercially available contrast agents as detailed in Supplementary Note 1.

## Native-state imaging at ESRF ID17 beamline

C57BL/6J mice (strain code 632; Charles River Laboratories, France) were used for all experiments at the ID17 beamline. To ensure proper acclimatization, animals were housed within the local animal facility for at least one week ahead of the experiments. All procedures complied with the European Directive 2010/63/EU on the protection of animals used for scientific purposes. Experimental protocols were reviewed and approved by Comité d'éthique en expérimentation animale de l'ESRF (ETHAX), under approval number APAFIS #30913-2021040211343677 v1.

For the dataset shown, a 12-week-old male mouse (SubjectID: Mouse19; body weight: 26.9 g) was premedicated with a subcutaneous injection of buprenorphine (0.1 mg/kg body weight) for analgesia. Anesthesia was induced 30 min later, after onset of analgesia, via intraperitoneal injection of a ketamine (73 mg/kg) and medetomidine (0.18 mg/kg) cocktail. During induction, the animal was placed in a custom-built heated chamber maintained at 33 °C. Anesthetic depth was monitored via reflex testing, and supplemental doses were administered as needed. To prevent corneal drying, eye ointment was applied, and to reduce potential X-ray imaging artifacts, the skull, neck, and upper thoracic region were shaved. Two separate subcutaneous injections of 0.5 ml glucose 10% were administered for fluid support. Tracheotomy surgery was then performed, and a metal tracheal cannula with Y-adapter (outer diameter: 1.0 mm, length: 13 mm, Hugo Sachs Elektronik, March-Hugstetten, Germany) was inserted to enable artificial ventilation.

The animal was then transferred to the imaging hutch containing the SRµCT setup described above and imaged using a monochromatic X-ray beam at a photon energy of 37.95 keV. Anesthesia was maintained throughout imaging via an intraperitoneal infusion line (30G needle with 0.28 mm inner diameter tubing) connected to a remote-controlled syringe pump. A total of 2 000 radiographs were acquired over a 180° rotation using a pco.edge 5.5 camera coupled to a Hasselblad 100 mm f/2.2 lens and a 250 µm LuAG:Ce scintillator, yielding an effective pixel size of 6.3 µm. The detector field of view was limited by the vertical extent of the beam, resulting in an effective field of view of 2560 × 780 pixels (16.1 mm × 4.9 mm). Images were acquired with an exposure time of 5 ms and an inter-projection overhead of 5 ms, resulting in a total scan time of 20 s. The sample-to-detector distance was 3 m.

Tomographic reconstruction was performed using GPU-accelerated filtered backprojection as implemented in the ASTRA Toolbox (version 2.1.0)[76,77], accessed through TomoPy (version 1.12.2)[78,79]. Propagation-based phase retrieval according to Paganin et al. was applied, using a δ/β parameter of 200[50].

## Contrast-enhanced imaging at SPring-8 BL20B2 beamline

C57BL/6JJmsSLC mice (Japan SLC, Inc.) were used for all experiments at the BL20B2 beamline. To ensure proper acclimatization, animals were housed in the local animal facility and habituated to non-aversive handling techniques for 10 days prior to the experiments. The experimental protocols were reviewed and approved by the responsible ethics committee of SPring-8.

For the dataset shown, a 12-week-old female mouse (SubjectID: JP28; body weight: 20.2 g) was anesthetized with an intraperitoneal injection of medetomidine (0.3 mg/kg), midazolam (4 mg/kg), and butorphanol (5 mg/kg). To prevent hypothermia, the entire surgical area and imaging hutch were heated to 32 °C. Anesthetic depth was monitored via reflex testing, and supplemental doses were administered as needed. Eye ointment was applied, and the skull, neck, and upper thoracic region were shaved. Two separate subcutaneous injections of 0.5 ml glucose 10% were administered for fluid support. Tracheotomy was then performed, and a metal tracheal cannula with Y-adapter was inserted to enable artificial ventilation.

The mouse was then secured in a stereotactic frame and ventilated with a MiniVent ventilator (Model 845, Hugo Sachs Elektronik, March-Hugstetten, Germany), delivering 96% oxygen from an oxygen accumulator and 0.5-2% isoflurane. Exhaled air was passed through a water column to maintain a positive end-expiratory pressure of 2 cmH$_2$O. Ventilator settings included a stroke volume of 125 µl and a respiratory rate of 150 breaths per minute. Buprenorphine (0.1 mg/kg) was administered subcutaneously for prolonged analgesia only after induction with isoflurane. This was done to avoid antagonism with butorphanol administered earlier as part of the anesthesia protocol.

The ventricular infusion system consisted of plastic tubing with an outer diameter of 1.14 mm and an inner diameter of 0.69 mm. One end of the tubing was connected to a 25 µl Hamilton syringe via a 1 mm compression fitting. The other end was glued to a 28G PEEK microcannula of 2.3 mm length (32OP/PK/SPC, Protech International Inc., Boerne, United States). Infusion was performed using a syringe pump. Prior to cannula implantation, the syringe was mounted in the pump and, together with the connected tubing, filled with water to serve as hydraulic fluid. Care was taken to ensure that no air bubbles were present in the system. To prevent mixing at the interface, 2 µl of air was aspirated between the hydraulic fluid and the contrast agent. The contrast agent, diluted to a concentration of 240 mg Ba/ml, was drawn into the system immediately prior to cannula implantation to avoid drying or the introduction of air bubbles.

For cannula implantation, excess periosteum from the bone was removed and the bregma was identified. A small hole of about 1 mm diameter was drilled through the parietal bone at the injection coordinates. The microcannula was carefully inserted into the right lateral ventricle (Fig. 4a) using the stereotactic frame, following established protocols[7,80]. Medio-lateral and rostral-caudal coordinates for insertion were adjusted to account for anatomic differences in local mouse strains and inter-surgeon variability. For this procedure, the final coordinates were 1.2 mm lateral and 0.22 mm caudal to bregma.

The animal was then removed from the stereotactic frame and Minivent ventilator, and transferred to the imaging hutch containing the SRµCT setup, mounted in the holder, and ventilated with the SAR-1000 ventilator as described above. Anesthesia was maintained with 0.5 - 2% isoflurane, according to vital signs. Imaging was performed using a monochromatic X-ray beam at a photon energy of 40 keV, generated by a double multilayer monochromator with 4.8% energy bandwidth. Contrast agent infusion was initiated with the flow rate ramped linearly from 0 to 0.25 µl/min over the first minute and maintained at that rate until 5 µl of contrast agent had been delivered. To prevent CSF backflow into the infusion cannula, a very low maintenance flow rate of 0.02 µl/min was applied thereafter. The dataset shown in Fig. 2b was acquired 40 min after infusion start, at which point a total of 5.39 µl of contrast agent had been administered.

A total of 1800 radiographs were acquired over a 180° rotation using a Hamamatsu Orca Flash4.0 v2 camera coupled to a tandem lens system (105 mm f/2.4 and 85 mm f/1.4) and a 500 µm LuAG:Ce scintillator, yielding an effective pixel size of 8.0 µm. The detector field of view was limited by the vertical extent of the beam, resulting in an effective field of view of 2048 × 1500 pixels (16.4 mm × 12.0 mm). Images were acquired with an exposure time of 5 ms and an inter-projection overhead of 8 ms, resulting in a total scan time of 23 s. The sample-to-detector distance was 1 m. Tomograms were reconstructed using the BL20B2 in-house reconstruction software, ct-rec[81] (version 2023.05.03), with automatic center-of-rotation determination. Output image intensity values directly correspond to linear attenuation coefficients.

To correlate contrast agent concentrations with the linear attenuation coefficients, calibration scans were performed using a series of contrast agent dilutions. These were prepared by diluting a 1.5x concentrated suspension of the commercial product (480 mg Ba/ml, custom-produced by the manufacturer) with isotonic mannitol solution (54.65 mg/ml, the same suspension medium as used by the manufacturer), yielding final concentrations of 48, 120, 158.4, and 240 mg Ba/ml. At each of these concentrations, as well as at zero concentration, 4 µl of solution was drawn into a 0.69 mm inner diameter PVC tube using a syringe pump. These calibration samples were then imaged using the same acquisition settings as described above. The solution volumes were segmented by manual contouring in Python using the morphological_geodesic_active_contour algorithm from the scikit-image package. This process enabled the determination of mean gray values, which correspond to the mean X-ray linear attenuation coefficients.

To compare the contrast-to-noise ratio between native-state and contrast-enhanced imaging in the same sample, a pre-infusion scan performed 20 min prior to contrast administration was compared to an acquisition 40 min after the start of the infusion. A ROI measuring 30×30 voxels in-plane centered in the dorsal third ventricle and extending 170 voxels in depth along the anteroposterior axis was chosen to determine the mean X-ray linear attenuation coefficient ($\mu_1$) and its standard deviation ($\sigma_1$) in CSF. A second ROI of identical dimensions was placed in CSF-free parenchyma dorsal to the third ventricle to obtain corresponding values $\mu_2$ and $\sigma_2$. The contrast-to-noise ratio was calculated as $CNR = \frac{|\mu_1 - \mu_2|}{\sqrt{\sigma_1^2 + \sigma_2^2}}$.

## Native-state imaging of peri-mortem morphological changes at SPring-8 BL20B2 beamline

For the dataset shown, a 12-week-old female mouse (SubjectID: JP34; body weight: 19.2 g) was anesthetized with an intraperitoneal injection of a cocktail of medetomidine (0.3 mg/kg), midazolam (4 mg/kg), and butorphanol (5 mg/kg). The animal was then shaved and transferred to the imaging hutch without further surgical intervention. Immediately prior to imaging – 3 min before scan start –, euthanasia was induced by administering a five-fold overdose of the same anesthetic cocktail.

Imaging was performed as described above for SubjectID JP28, with two modifications: the sample-to-detector distance was increased to 3 m, and the effective voxel size was slightly smaller at 7.92 µm. To prevent oversaturation from edge enhancement, exposure time was reduced from 5 ms to 4 ms. As the camera frame rate remained unchanged, the total scan time per rotation remained 23 s. The rotation stage reset to its start position in 9 s, and the next scan started after an additional 8 s, resulting in one scan every 40 s. In total, 60 time points were recorded over a period of 40 min.

To isolate potential ventricular motion from whole-animal displacement, reconstructions from each time point were rigidly registered to the first reconstruction in the series, referred to as the reference image. To prevent contrast-enhanced ventricular regions from influencing the registration, an extended bone mask excluding the ventricles was generated for the reference image. Registration was primarily driven by maximizing normalized cross-correlation with this mask as the image similarity metric. For time points 28, 29, and 60, mutual information was used instead, as registration with cross-correlation failed to yield satisfactory alignment (correlation coefficient below 0.7). Automatic image registration was performed using elastix software (version 4.9)[82,83] on images with reduced bit depth to reduce memory footprint and increase processing speed. The resulting transformation parameters were then applied to the original images with full dynamic range.

## Contrast-enhanced imaging of peri-mortem morphological changes at CLS BMIT beamline

C57BL/6J mice (stock 000664; Jackson Laboratories) were used for all experiments at the BMIT beamline. To ensure proper acclimatization, the animals were housed in the local animal facility for at least one week prior to the experiments. The experimental protocols were reviewed and approved by the University of Saskatchewan Animal Research Ethics Board (approval number AUP 20230072 MOD#1).

For the dataset shown, a 12-week-old male mouse (SubjectID: CA019; body weight: 28.5 g) was premedicated with a subcutaneous injection of buprenorphine (0.1 mg/kg) for analgesia. Anesthesia was induced 30 min later, following onset of analgesia, via intraperitoneal injection of a ketamine (50 mg/kg) and medetomidine (1 mg/kg) cocktail. The animal was shaved and kept warm using heating pads. Two subcutaneous injections of 0.5 ml 10% glucose solution were administered for fluid support, and tracheotomy was performed using a custom 3D-printed, radiotransparent plastic tracheal cannula with Y-adapter.

The mouse was then secured in a stereotactic frame and ventilated with a MiniVent, delivering 96% oxygen from an oxygen accumulator and 0.5-2% isoflurane. The exhaled air was routed through a water column to maintain a positive end-expiratory pressure of 2 $cmH_2O$. Stroke volume was set to 150 µl, and respiration rate to 140 breaths per minute. A 28G PEEK infusion cannula was filled with contrast agent at standard concentration (320 mg Ba/ml) and implanted into the right lateral ventricle as described above. Cannula coordinates were 1.2 mm lateral and 0.03 mm caudal to the bregma.

The animal was then transferred to the imaging hutch[84] containing the SRµCT setup, maintained under 0.5-2% isoflurane anesthesia with the SAR-1000 ventilator, and imaged at a photon energy of 37.8 keV, generated by the beamline's double bent Laue crystal monochromator. A total of 2 000 radiographs were acquired over a 180° rotation using a pco.edge 4.2 CLHS camera coupled to a tandem lens system (105 mm f/2.4 and 85 mm f/1.4) and a 200 µm LuAG:Ce scintillator, yielding an effective pixel size of 8 µm. The detector field of view was limited by the beam dimensions, resulting in an effective field of view of 2048 × 1200 pixels (16.4 mm × 9.6 mm). Images were acquired with an exposure time of 15 ms and an inter-projection overhead of 1 ms, resulting in a total scan time of 32 s. The sample-to-detector distance was 0.6 m, and the source-to-sample distance was 57.8 m.

The *in vivo* scan was initiated 58 min after the start of contrast agent infusion. The flow rate was linearly increased from 0 to 0.25 µl/min over the first minute and kept at that rate until a total of 5 µl had been delivered, after which it was reduced to the maintenance infusion rate of 0.02 µl/min to prevent backflow. By the time of *in vivo* imaging, an overall volume of 5.75 µl had been infused. Infusion was terminated 60 min after initiation, and the animal was euthanized via intraperitoneal injection of sodium pentobarbital (300 µl, 240 mg/ml) at 74 min post-infusion start. Time of death was determined as 78 min, defined by a drop in peak end-tidal $CO_2$ to 0.1%, which served as the most reproducible physiologic marker. The post-mortem scan was performed 79 min after infusion start, corresponding to 1 min after confirmed death.

Tomograms were reconstructed using the image processing toolkit tofu, which utilizes the UFO framework as its back-end[85]. The center of rotation was determined automatically. Output image intensity values correspond to $\mu\Delta x$, the product of the linear attenuation coefficient ($\mu$) and the voxel length ($\Delta x$).

The resulting 3D datasets were rigidly registered to the first time point and transformed as described above. All datasets were successfully registered using normalized cross-correlation; registration based on mutual information was not required. Segmentation of the ventricular spaces for 3D visualization was performed in Amira 3D (version 2023.2, Thermo Fisher Scientific, Waltham, United States). Segmentation parameters were selected based on visual inspection to yield the best results in each case. For the *in vivo* dataset, a 4.5-pixel-radius spherical median filter was applied for denoising, followed by manual thresholding to generate a binary mask for bones ($M_{bones}$). To remove ventricular edge regions that were erroneously included in the mask, connected components (6-voxel connectivity) were color-coded and manually excluded after visual inspection. The final mask was dilated using a large spherical kernel with a 12.5-pixel radius to ensure full coverage of bones.

To calculate ventricular volumes in live and post-mortem states, the masks $M_{live}$ and $M_{post-mortem}$ were generated by automatic thresholding of the non-denoised and median-filtered images, respectively. For $M_{post-mortem}$, a median filter with a 1.5-pixel radius spherical kernel was applied prior to thresholding. In both cases, thresholding was performed using Otsu's method[86]. Bones were excluded by applying the mask $M_{bones}$ using a logical AND NOT operation. Connected component analysis (6-voxel connectivity) with manual component selection was then performed, followed by morphological closing with a 7.5-pixel-radius spherical kernel to fill in the unenhanced ChP located within the ventricles. Volume estimates for the ventricular masks were calculated using the material statistics module in Amira 3D, based on the number of voxels multiplied by the voxel volume. Cross-sectional areas in each coronal slice were obtained using the area module, which calculates the number of non-zero pixels multiplied by the pixel area.

A dedicated local segmentation was used to quantify peri-mortem changes in the narrow spaces of the aqueduct (Fig. 3a-d). Manual segmentation was performed on non-denoised images, followed by morphological closing with a 1-pixel-radius spherical kernel and morphological hole filling for regions up to 64 voxels in size. The largest connected component (26-voxel connectivity) was then extracted and verified by visual inspection. These operations were implemented in a Python pipeline (version 3.13.1) using the scikit-image library (version 0.25.0)[87]. Based on the resulting segmentations, volumes, surface areas, cross-sectional areas, and hydraulic diameters were calculated. For plotting, cross-sectional area and hydraulic diameter profiles were smoothed with a Savitzky-Golay filter (window length: 20, polynomial order: 2; SciPy, version 1.15.1)[88]. For 3D visualization, surface meshes of the aqueduct were generated using PyVista (version 0.44.2)[89] and smoothed with a Laplacian filter (1500 iterations, relaxation factor: 0.01) to reduce pixelation artifacts.

## Mapping spatiotemporal solute distribution throughout the cranial CSF space

### Intra-cerebroventricular infusion at ESRF ID17 beamline

For the dataset shown, a 12-week-old female mouse (SubjectID: Mouse63; body weight: 22.8 g) was anesthetized and implanted with a contrast agent infusion cannula as described above for the previous ESRF ID17 experiment (SubjectID: Mouse19), with minor modifications. The animal was artificially ventilated using a MiniVent and oxygen-enriched air during surgery. Exhaled air was routed through a water column to maintain a positive end-expiratory pressure of 2 cmH$_2$O. Ventilator settings included a stroke volume of 125 µl and a respiratory rate of 150 breaths per minute. A 2.3 mm PEEK infusion cannula was filled with 1.5× concentrated contrast agent (480 mg Ba/ml) and implanted into the right lateral ventricle. Cannula coordinates were 0.95 mm lateral and 0.22 mm caudal to the bregma.

The animal was then transferred to the imaging hutch and mounted in the SRµCT setup as described above. Imaging was performed using a monochromatic X-ray beam at a photon energy of 37.5 keV. Contrast agent was infused while tomographic scans were acquired every 30 s. During the first 5 min, 1 µl of contrast agent was infused at a rate of 0.2 µl/min, after which the infusion was stopped. For each scan, 2000 radiographs were acquired over a 360° rotation using a pco.edge 5.5 camera coupled to a Hasselblad 100 mm f/2.2 lens and a 250 µm LuAG:Ce scintillator, yielding an effective pixel size of 6.45 µm. Due to the limited vertical extent of the beam, the detector field of view was restricted to 2560 × 700 pixels. Images were acquired with an exposure time of 4 ms and an inter-projection overhead of 1 ms, yielding a total scan time of 10 s per time point. The complete time series consisted of 100 scans acquired over a total duration of 50 min. Flat-field images were acquired only at the beginning and end of the series. The sample-to-detector distance was 2.5 m.

Tomograms were reconstructed using the ESRF software Nabu (version 2023.2.0), which implements a filtered backprojection algorithm[90]. Ring artifacts were mitigated by stripe removal in the sinograms using a combined wavelet-Fourier filtering approach[91]. The center of rotation for each reconstruction was manually fine-tuned. All tomograms were rigidly registered to the first time point using normalized cross-correlation, as described above.

Spatial resolution was estimated as the inverse of the first crossover point of the Fourier shell correlation (FSC) curve with the 1-bit threshold line (Fig. 4e), which was calculated using a Fourier-space signal-to-noise ratio of 0.5 to account for the use of two half-datasets[53]. The FSC was computed on a 256×470×470 voxel sub-volume containing both bone and soft tissue. The two tomograms, each reconstructed from half the set of projections, were multiplied with a Hamming window prior to applying the discrete Fourier transform[57]. The resulting FSC curve was smoothed using a third-order Savitzky-Golay filter with a window width of 50[88].

## Intra-cisterna magna infusion at ESRF ID17 beamline

For the dataset shown, a male mouse (SubjectID: Mouse50) twelve weeks of age and 23.7 g body weight was anesthetized and shaved as in the other ESRF ID17 experiments described above.

The cisternal infusion system consists of a 30G needle connected via tubing to a 1 ml syringe, following the procedure described by Xavier et al.[92], with minor modification. The needle was carefully cut using a small rotary saw, instead of breaking off the needle with a bevel of an insulin syringe. To ensure removal of any metal residues from the sawing process, the needle was rinsed by submerging it in water and then thoroughly flushed. Before the start of infusion implant surgery, syringes and tubing were mounted into the syringe pump and filled with mineral oil as a hydraulic fluid, ensuring that no air bubbles were introduced. To prevent the mixing of fluids, 2 μl of air was drawn between the hydraulic fluid and the contrast agent, preventing a mixed fluid interface. 1.5× concentrated contrast agent (480 mg Ba/ml) was then drawn into the system shortly before implantation, to avoid drying or the introduction of air bubbles.

The mouse was secured in a stereotactic frame and a small skin incision was made over the occipital bone. The three muscle layers covering the cisterna magna were carefully dissected under a stereomicroscope using fine forceps and scissors. The atlantooccipital membrane was then perforated, and the cannula needle was gently inserted into the cisterna magna (Fig. 4a). A metal clip was used to stabilize the cannula during insertion and was gently removed once the cannula was securely in place and sealed with tissue glue.

The animal was then transferred to the radiation hutch and imaged using the same SRμCT setup and parameters as Mouse63 described above. Tomograms were reconstructed and rigidly registered using the same protocol, as well.

The animal was infused with contrast agent while one tomographic scan was acquired every 30 seconds for 25 min. 2.5 μl of contrast agent were infused at rate of 0.5 μl/min in the first 5 min of the time series, then infusion was stopped. Flat-field images were only acquired at the beginning and the end of the 50 scans.

## Quantifying tissue motion

### Retrospective cardiac-gated imaging at ESRF ID17 beamline

For the dataset shown, a 12-week-old male mouse (SubjectID: Mouse17; body weight: 23.7 g) was anesthetized and shaved as described above for the other ESRF ID17 experiments. The animal was secured in a stereotactic frame without artificial ventilation. A total of 5 μl of contrast agent was injected into the lateral ventricle at a flow rate of 0.5 μl/min using a 34G Hamilton syringe. Injection coordinates were 0.95 mm lateral and 0.22 mm caudal to the bregma, at a depth of 2.3 mm.

The animal was then transferred to the imaging hutch and mounted in the SRμCT setup as described above. Imaging was performed using a monochromatic beam at a photon energy of 37.95 keV. A total of 60 000 radiographs were acquired over a 360° rotation using a pco.edge 5.5 camera coupled to a Hasselblad 100 mm f/2.2 lens and a 250 μm LuAG:Ce scintillator, yielding an effective pixel size of 6.3 μm. Due to the limited vertical extent of the beam, the detector field of view was restricted to 2560 × 780 pixels. Images were acquired with an exposure time of 5 ms and an inter-projection overhead of 5 ms, resulting in a total scan time of 10 min. The sample-to-detector distance was 3 m.

Retrospective cardiac gating was performed using MATLAB (release R2022b; The MathWorks Inc., Natick, United States) following the approach described by Fardin et al.[93]. The ECG signal was denoised using a discrete wavelet transform. R-peaks were detected using the findpeaks function, with the parameters minimum peak height and minimum peak *distance* optimized via grid search to minimize the error relative to the recorded heart rate. Following peak detection, the time delay between the recorded trigger signal of each acquired projection and the closest R-peak was calculated. Projections were then grouped into time bins of 10 ms width based on their time delay, resulting 18 bins within the minimum cycle duration and 2 188 projections per bin. For each bin, a tomogram representing a distinct phase of the cardiac cycle was reconstructed using GPU-accelerated filtered backprojection, implemented in the ASTRA Toolbox (version 2.1.0)[76,77], and accessed through TomoPy (version 1.12.2)[78,79].

For initial analysis in 2D, images corresponding to the most decorrelated cardiac phases, i.e., 10 ms and 150 ms after the R-peak, were subtracted to visually identify deformations of brain structures (Fig. 5c, h, j). Structures exhibiting intensity difference exceeding the noise floor were identified as having moved between the two phases. This inspection was conducted on both raw and median-filtered images.

To quantify the detected motion in 3D, a ROI encompassing the nasopharynx was cropped from two scans that were cardiac-gated at 10 ms and 150 ms after the cardiac R-peak. Segmentation of the nasopharynx was performed using a coarse mask to isolate it from adjacent bone structures, applying a combination of thresholding, morphological operations, and connected component analysis. Specifically, the nasopharyngeal perimeter was delineated through the following steps: binarization with a manually selected threshold; morphological closing followed by opening with a 2-pixel-radius spherical kernel; removal of connected regions with volumes below 512 voxels (to eliminate noise); further closing with a 1.5-pixel-radius kernel; extraction of the largest component using full connectivity; and final morphological closing with a 2.5-pixel-radius kernel. The resulting perimeter was filled using flood-filling to generate a closed mask.

All processing steps were implemented in a Python pipeline (version 3.13.1) using the scikit-image library (version 0.25.0)[87]. Nasopharyngeal surface meshes were generated from the segmentations using PyVista (version 0.44.2)[89] and smoothed using a Laplacian filter (1 000 iterations, relaxation factor: 0.01) to reduce pixelation artifacts (Fig. 5e). Surface-to-surface distances were calculated for each mesh point in the 150 ms dataset by determining the distance to the nearest-neighbor point in the 10 ms mesh through KDtree in SciPy (version 1.15.1) (Fig. 5g).

## Imaging choroid plexus motion at CLS BMIT

The dataset shown was acquired from the same animal (SubjectID: CA019) as used in the peri-mortem morphological analysis at CLS BMIT, but corresponds to a separate imaging series. Animal procedures, experimental setup, image reconstruction and rigid registration protocols were identical.

One tomographic dataset was acquired every minute over a 50 min period during continuous contrast agent infusion. Each scan lasted 32 s, followed by an 18 s return of the rotation stage to its initial position, and a 10 s delay before the next scan. The contrast agent flow rate was ramped linearly from 0 to 0.25 µl/min during the first minute, maintained at this rate until a total volume of 5 µl had been delivered, and then reduced to 0.02 µl/min for the remainder of the experiment. Flat-field images were acquired only at the beginning and end of the 50-scan series.

To visualize motion over short time intervals, subtraction images were generated in coronal and transverse planes between three consecutive time points – 36, 37, and 38 min after the start of infusion – corresponding to 1 min and 2 min intervals, respectively (Fig. 5i,k).

The 3D quantification of choroid plexus motion was based on datasets acquired 2 min apart (at 36 and 38 min after infusion start). The choroid plexuses of the right and left lateral ventricles were segmented using a combination of thresholding, morphological operations, and connected component analysis. All processing steps were performed on the left and right sides independently.

First, ROIs were cropped from both datasets. Ventricular segmentation was performed by binarizing the volumes using a threshold determined via Otsu's method[86], followed by morphological closing with a spherical kernel of 8-pixel radius, removal of holes smaller than 32 voxels, extraction of the largest connected component, and morphological erosion using a 5-pixel-radius spherical kernel. Within this ventricular mask, the choroid plexus was segmented via binarization (again using Otsu's method), removal of holes smaller than 4 voxels, and extraction of the largest connected component (26-voxel connectivity). All operations were implemented in a Python pipeline (version 3.13.1) using the scikit-image library (version 0.25.0)[87]. Surface meshes of the choroid plexus were generated using PyVista (version 0.44.2)[89] and smoothed with a Laplacian filter (1 000 iterations, relaxation factor: 0.01) to reduce pixelation artifacts (Fig. 5l). To estimate motion magnitude, correspondence between the surface points of the meshes from the two time points was established through iterative closest point registration using the Open3D library (version 0.18.0)[94] and nearest-neighbor analysis through a KDtree in SciPy (version 1.15.1). The distances between corresponding points were then calculated on the non-registered meshes (Fig. 5l,m).

# Data availability

Reconstructed 3D SRµCT datasets are available under a Creative Commons Attribution 4.0 International license via individual dataset entries on Zenodo, linked through the main Zenodo repository[49]. For time series experiments, a single reconstructed 2D section is provided for each time point. Full 3D stacks are included for only one or two representative time points due to their cumulative size of 2.7 terabytes. The remaining datasets are available upon request. Computer-aided design (CAD) files of the mouse holder, infusion cannula, intra-cisterna magna infusion and tracheotomy stage, along with extended metadata and quantitative results tables, are also available in the main Zenodo repository. Information on other available datasets can be accessed via the FABRIC4 portal[70].

# Code availability

Code is available on Zenodo[49] under GNU General Public License v3.0 or later. The repository includes Python scripts for calculating Fourier shell correlation (FSC) and Fourier ring correlation (FRC), MATLAB and elastix code for image registration, Python code using scikit-image along with Amira project files for image segmentation, and Python code using PyVista to generate meshes from the segmentations. It also includes MATLAB code for retrospective cardiac gating, LabVIEW code for animal monitoring equipment, Arduino code for ventilator synchronization, as well as Nabu scripts and Python code using TomoPy for tomographic reconstruction.

## Acknowledgements

We acknowledge the ESRF for the provision of synchrotron radiation facilities under proposal numbers md1230 and md1324, and thank Michael Krisch for his assistance and support with beamline ID17. We also acknowledge SPring-8 for granting beamtime at beamline BL20B2 under proposal 2023A1208. Part of this research was performed under beamtime proposal 38G13327 at the BMIT beamline of the Canadian Light Source (CLS), a national research facility of the University of Saskatchewan, which is supported by the Canada Foundation for Innovation (CFI), the Natural Sciences and Engineering Research Council (NSERC), the National Research Council (NRC), the Canadian Institutes of Health Research (CIHR), the Government of Saskatchewan, and the University of Saskatchewan. We thank Fabian Eggiman and Daniel Junker of TPF AMF at the University of Zurich for their support with the design and manufacturing of the animal setup. This research was funded in part by the Swiss National Science Foundation (SNSF Sinergia 213535), the Fidelity Bermuda Foundation, and the National Plan for NRRP Complementary Investments (project number: PNC0000003 - Advanced Technologies for Human-Centred Medicine; project acronym: ANTHEM). For the purpose of open access, a CC BY 4.0 public copyright license is applied to any author-accepted manuscript (AAM) version arising from this submission.

## Ethics declarations

All procedures complied with the European Directive 2010/63/EU on the protection of animals used for scientific purposes. The experimental protocols used at ESRF (France) were reviewed and approved by the Comité d'éthique en expérimentation animale de l'ESRF (ETHAX), approval number APAFIS #30913-2021040211343677 v1. The experimental protocols used at SPring-8 (Japan) were reviewed and approved by the facility's responsible ethics committee. The experimental protocols used at CLS (Canada) were reviewed and approved by the Animal Research Ethics Board of the University of Saskatchewan, approval number AUP 20230072 MOD#1.

# Figure 1: Modular, portable setup enables in vivo imaging across multiple synchrotron radiation facilities

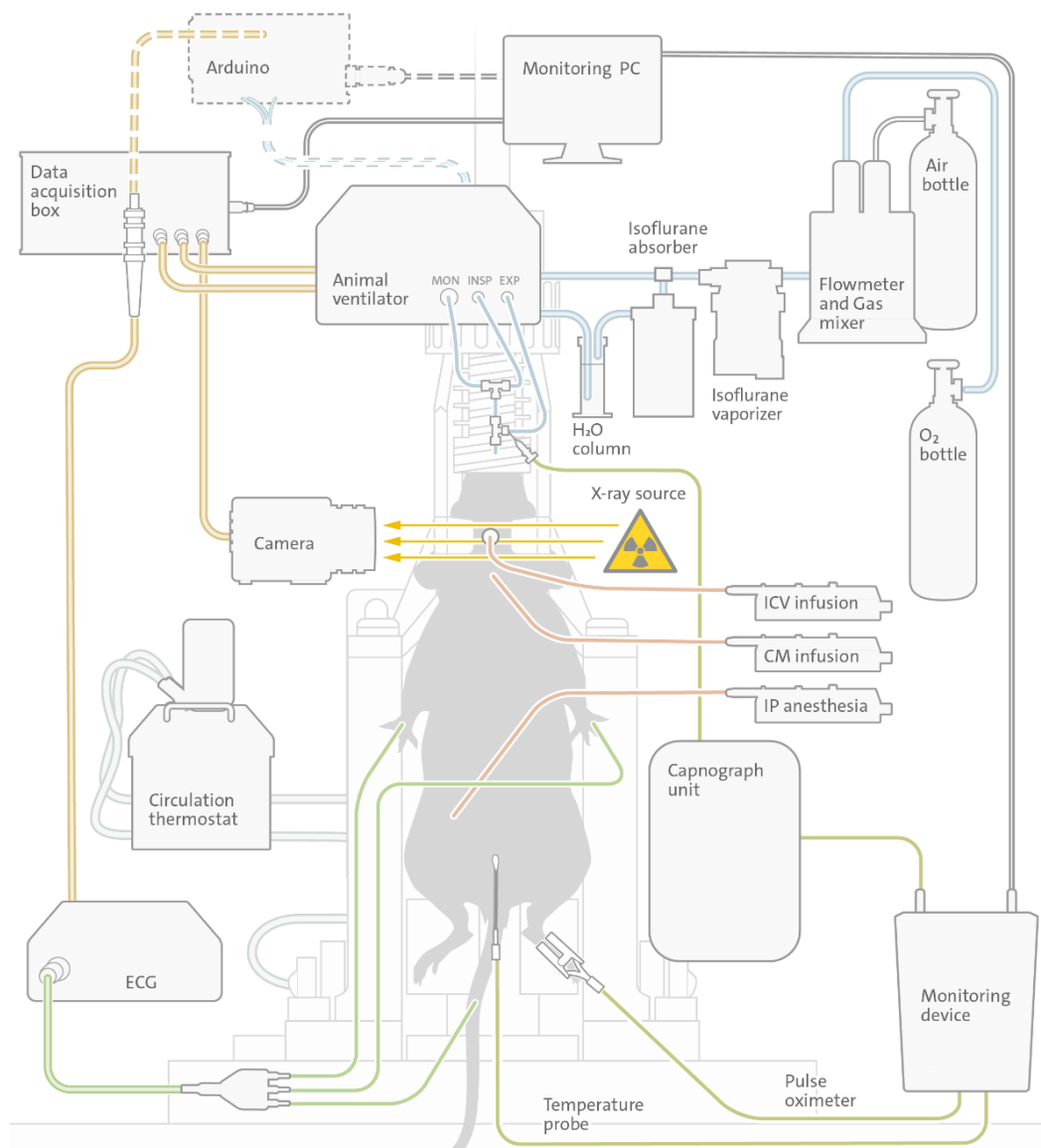

For *in vivo* SRμCT, mice were secured in a custom-designed animal holder. Body temperature was maintained using a circulation thermostat connected to the holder. Physiologic parameters, including temperature, blood oxygenation, and carbon dioxide partial pressure, were monitored using a PhysioSuite health monitoring device. Ventilation was provided by a SAR-1000 small animal ventilator. Dashed lines indicate optional components used for synchronization of cardiac and respiratory cycles via custom software running on an Arduino microcontroller. Positive end-expiratory pressure was maintained using a water column. An isoflurane absorber was used when gas anesthesia was administered. Electrocardiogram and ventilation signals were recorded through a USB data acquisition interface and monitored on a PC running either LabVIEW or LabChart software. Anesthesia options included tracheal administration of isoflurane mixed with pure oxygen, or intraperitoneal (IP) injection of anesthetics, in which case oxygen-enriched air was used for ventilation. A remote-controlled syringe pump was used to deliver intraperitoneal anesthesia. The same type of pump was also used for intra-cerebroventricular (ICV) and intra-cisterna magna (CM) infusion of contrast agent. Images were captured using pco.edge 5.5, pco.edge 4.2, or Orca Flash 4.0 v2 cameras. Figure by Tara von Grebel, University of Zurich, Information Technology, MELS/SIVIC, published under Creative Commons Attribution-NoDerivatives 4.0 International.

# Figure 2: Both contrast-enhanced and native-state imaging show cerebrospinal fluid spaces

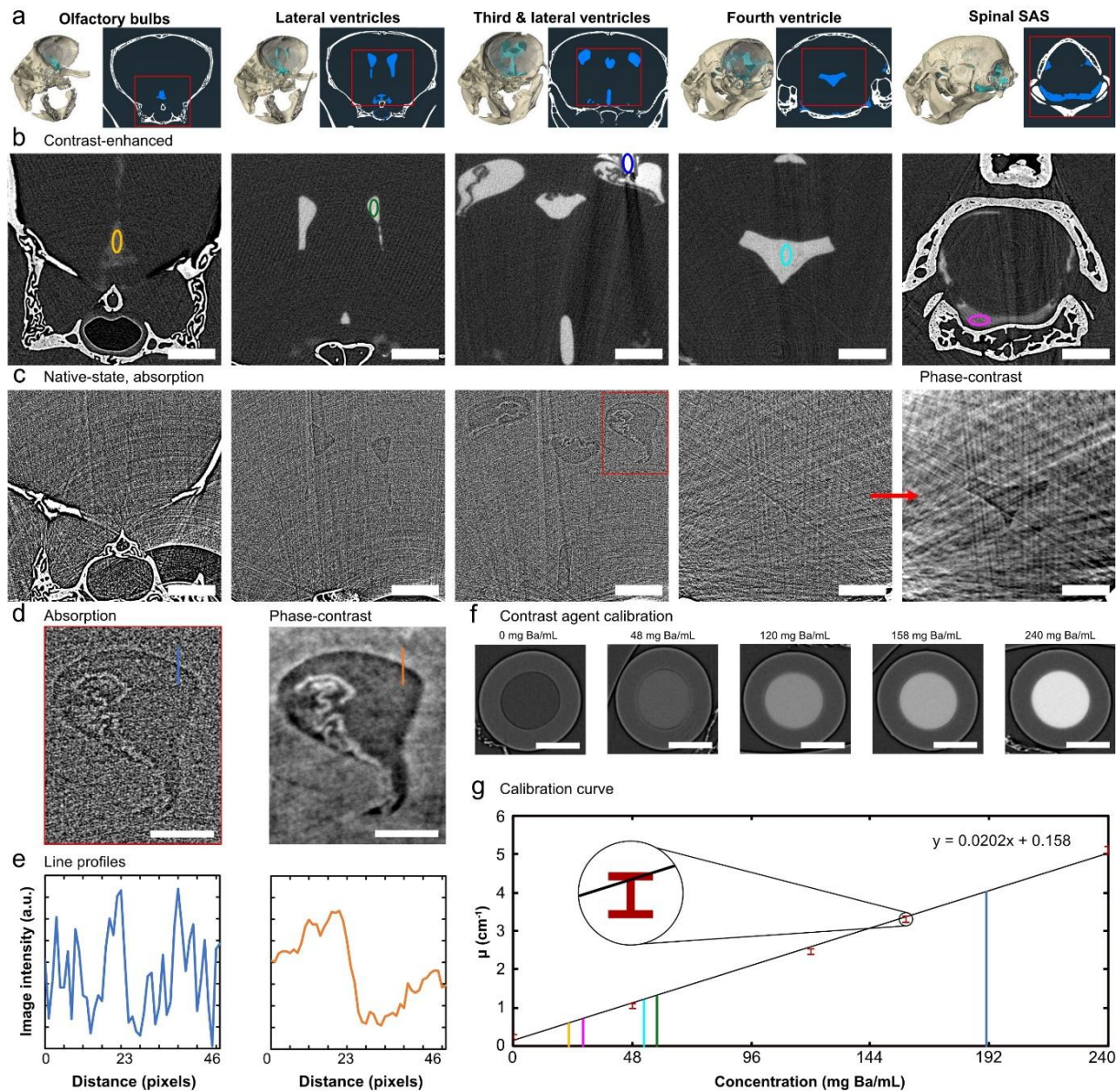

Contrast agent-free (native-state) tomography allows for CSF space imaging with minimal disturbance of fluid physiology, while contrast-enhanced scans improve fluid-tissue differentiation. **a**, Spatial orientation of the images shown in **b** and **c**. The 3D renderings (left in each column) show bone (beige) and CSF spaces (turquoise). Coronal cross-sections (right) correspond to similarly cut positions in the respective 3D rendering. The red outlined areas appear magnified in **b** and **c**. **b,** Magnified views of coronal sections through the reconstructed contrast-enhanced data framed red in **a**. Images were acquired 40 min after the start of ICV infusion. Attenuation coefficients within the framed regions were used to derive contrast agent concentration (**g**). Attenuation coefficient values in the range of [0.01 1.61] cm$^{-1}$ were mapped onto an 8-bit dynamic range for visualization in this figure. Scale bars: 1 mm. **c,** First four panels: magnified coronal views of areas marked in **a**, acquired in a different experiment and different mouse without contrast agent. Last panel: phase-retrieved reconstruction ($\delta/\beta$ = 200) of the data shown in the fourth panel. Figure dynamic range: [−1.03 1.39] cm$^{-1}$, scale bars: 1 mm. **d,** Comparison of fluid-tissue contrast before (left) and after (right) phase retrieval. The right lateral ventricle is shown, corresponding to the framed area in the center panel in **c**. Blue and orange bars indicate the positions of the line profiles in **e**. Scale bars: 0.5 mm. **e,** Profiles of image intensity across the ventricular wall along the lines shown in **d**. Grayscales values within the percentiles 1 to 99 were rescaled to [0 1]. **f,** Axial cross-sections of calibration tubes with increasing contrast agent concentration from left to right: 0, 48, 120, 158.4, and 240 mg Ba/ml. Figure dynamic range: [-1.5 6] cm$^{-1}$, scale bars: 0.5 mm. **g,** Linear fit of attenuation coefficient, μ, (vertical axis) to barium concentration, c, using the mean attenuation values of the inner region of the calibration samples in **f**, without first panel (no contrast agent), yielding $\mu = 0.0202$ ml/(mg·cm)·c + 0.158 cm$^{-1}$. Inset: Standard deviation of attenuation values in the 158.4 mg Ba/ml calibration sample. The vertical lines indicate the barium concentrations in the correspondingly color-coded regions in **b**.

# Figure 3: The ventricular CSF spaces contract peri-mortem

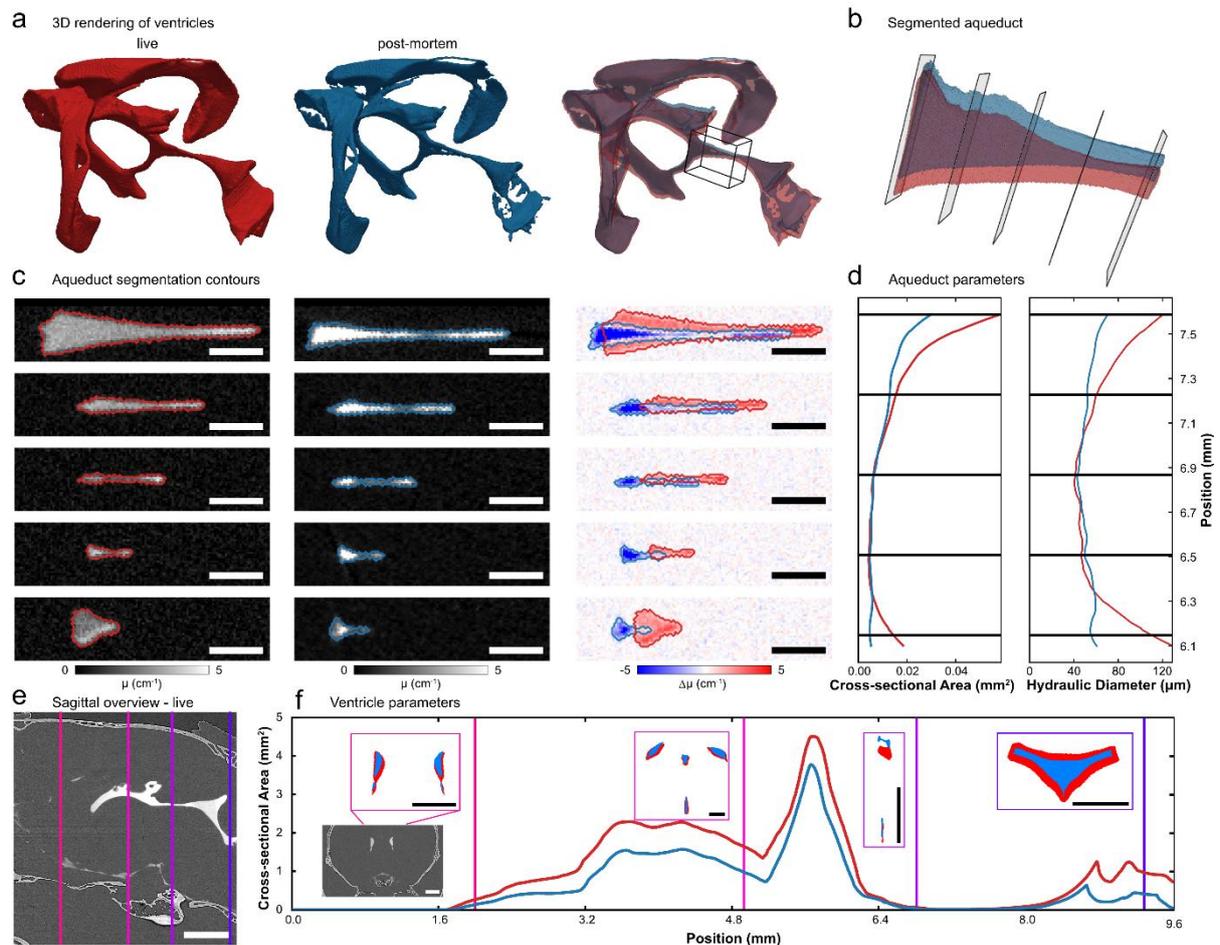

Quantification of the peri-mortem changes in cerebral ventricular volume and cross-section. **a**, Rendering of ventricles acquired *in vivo* (left, red, referred to as 'live'), immediately following cessation of vital signs 4 min after pentobarbital euthanasia (middle, blue, 'post-mortem'), and overlay of the two (right). The boxed region, magnified in **b**, contains the cerebral aqueduct. **b**, Rendering of semi-automatically segmented aqueduct and coronal section planes used in **c**. **c**, Contour lines delineating the aqueduct in the section planes shown in **b**. First column: *live*. Second column: *post-mortem*. Third column: *live minus post-mortem*. Scale bar: 200 μm. **d**, Cross-sectional area and hydraulic diameter of the aqueduct in live (red) and post-mortem (blue) states along the longitudinal axis of the box shown in **a**, with each data point representing one coronal cross-section. Vertical axis: distance from the end of the third ventricle (bottom, at 6.1 mm from the cribriform plate) to the beginning of the fourth ventricle (top, 7.6 mm). The black horizontal lines indicate the locations of the coronal planes shown in **b**. **e**, Sagittal view showing the location of the coronal planes used in **f**. Scale bar: 2 mm. **f**, Change in coronal cross-sectional area of the ventricles along the sagittal axis. Red line: *live*. Blue line: *post-mortem*. The starting position is the cribriform plate (0 mm mark on horizontal axis). Shapes of coronal ventricular cross-sections at the sagittal locations marked with colored vertical lines in **e** are shown in the insets adjacent to the vertical lines of the same respective color. Scale bars: 1 mm.

# Figure 4: Contrast agent infusion into lateral ventricle and cisterna magna show differences in local CSF dynamics

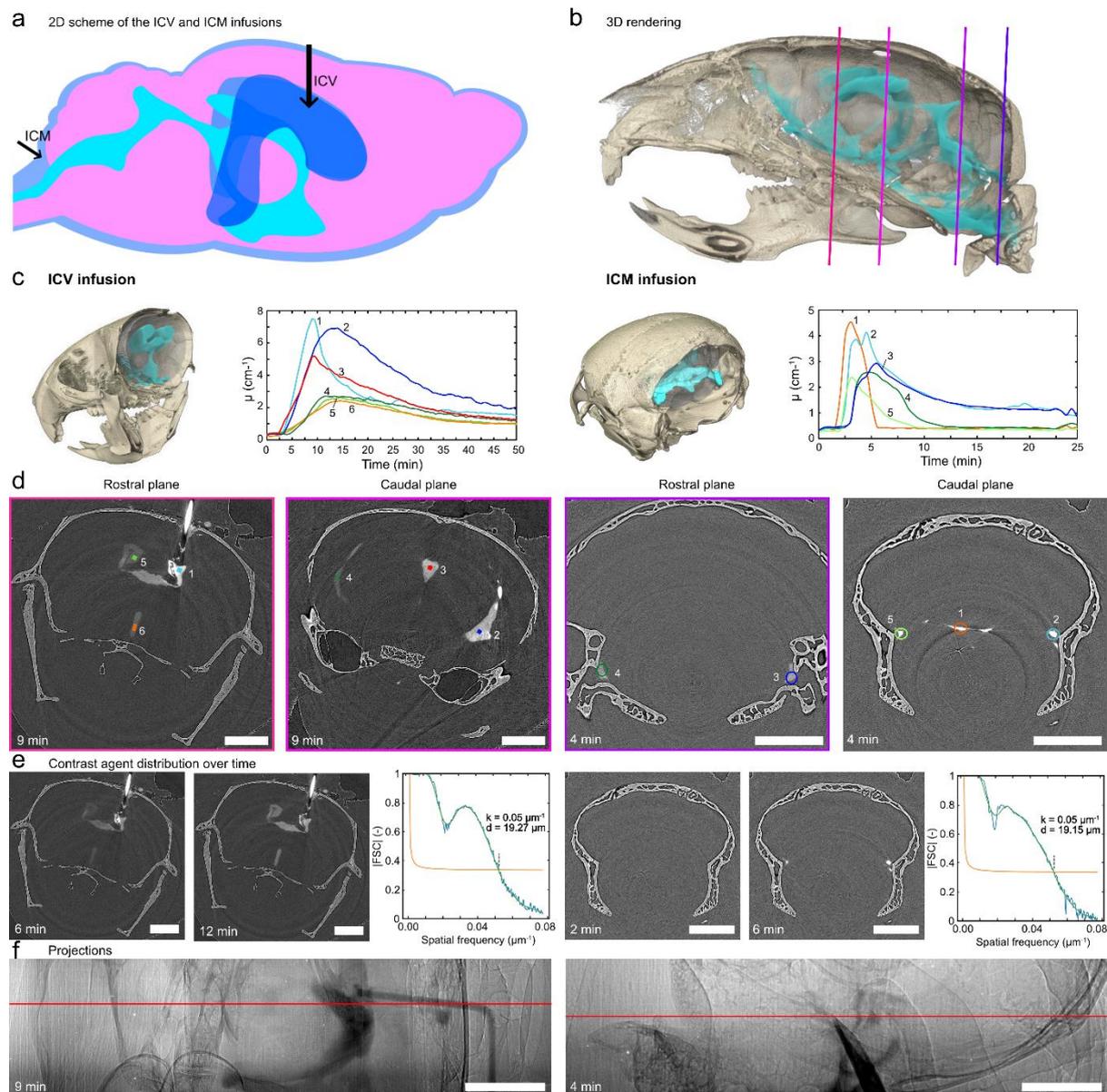

Spatiotemporal changes in contrast agent concentration upon intra-cerebroventricular (ICV) and intra-cisterna magna (ICM) infusions. **a,** Illustrative projection of CSF spaces (shades of blue) and brain (magenta) onto the sagittal plane for spatial orientation. Arrows indicate infusion locations. **b,** Rendering of half the skull with bone shown in beige and CSF spaces in turquoise. Lines indicate the orientation of image planes perpendicular to the sagittal plane in **d**. **c,** Change in contrast agent concentration, expressed as change in attenuation coefficient, over time upon ICV (left) and ICM (right) infusion for regions marked by the same color and number in the coronal cross-sections in **d**, namely: ICV, 1: right lateral ventricle, rostral, 2: right lateral ventricle, caudal, 3: third ventricle, caudal, 4: left lateral ventricle, caudal, 5: left lateral ventricle, rostral, 6: third ventricle, caudal. ICM, 1: mid cisterna magna, caudal, 2: subarachnoid space (SAS), right, caudal, 3: SAS, right, rostral, 4: SAS, left, rostral, 5: SAS, left, caudal. **d,** Contrast agent distribution after 9 min of ICV (left) and 4 min of ICM (right) infusion. Rostral and caudal coronal cross-sections are shown, positioned as displayed in **b**, spaced 1.9 mm and 1.6 mm apart, respectively. Scale bar: 2 mm. **e,** Temporal development of contrast agent distribution in, respectively, the rostral and caudal planes shown in **d** upon ICV (left) and ICM (right) infusion. Figure dynamic ranges were equally scaled for all time points from [-2.41 8.34] cm$^{-1}$ for ICV and [-3.37 6.33] cm$^{-1}$ for ICM infusion. Scale bar: 2 mm. Graphs display the raw (blue) and smoothed (green) Fourier shell correlation (FSC) curves used to estimate spatial resolution as limited by signal-to-noise, using the 1-bit threshold curve (yellow) for the ICV (left) and ICM (right) infusion, yielding resolutions d = 19.3 μm and d = 19.2 μm, respectively. **f,** Projections 9 and 4 min after the start of ICV (left) and ICM (right) infusion, chosen for the high contrast agent content in the field of view. Red lines indicate the location of the coronal cross-sections shown in **e**. Scale bar: 2 mm.

# Figure 5: Cardiac-gated and non-gated time series scans show movement of the nasopharynx and choroid plexus, respectively

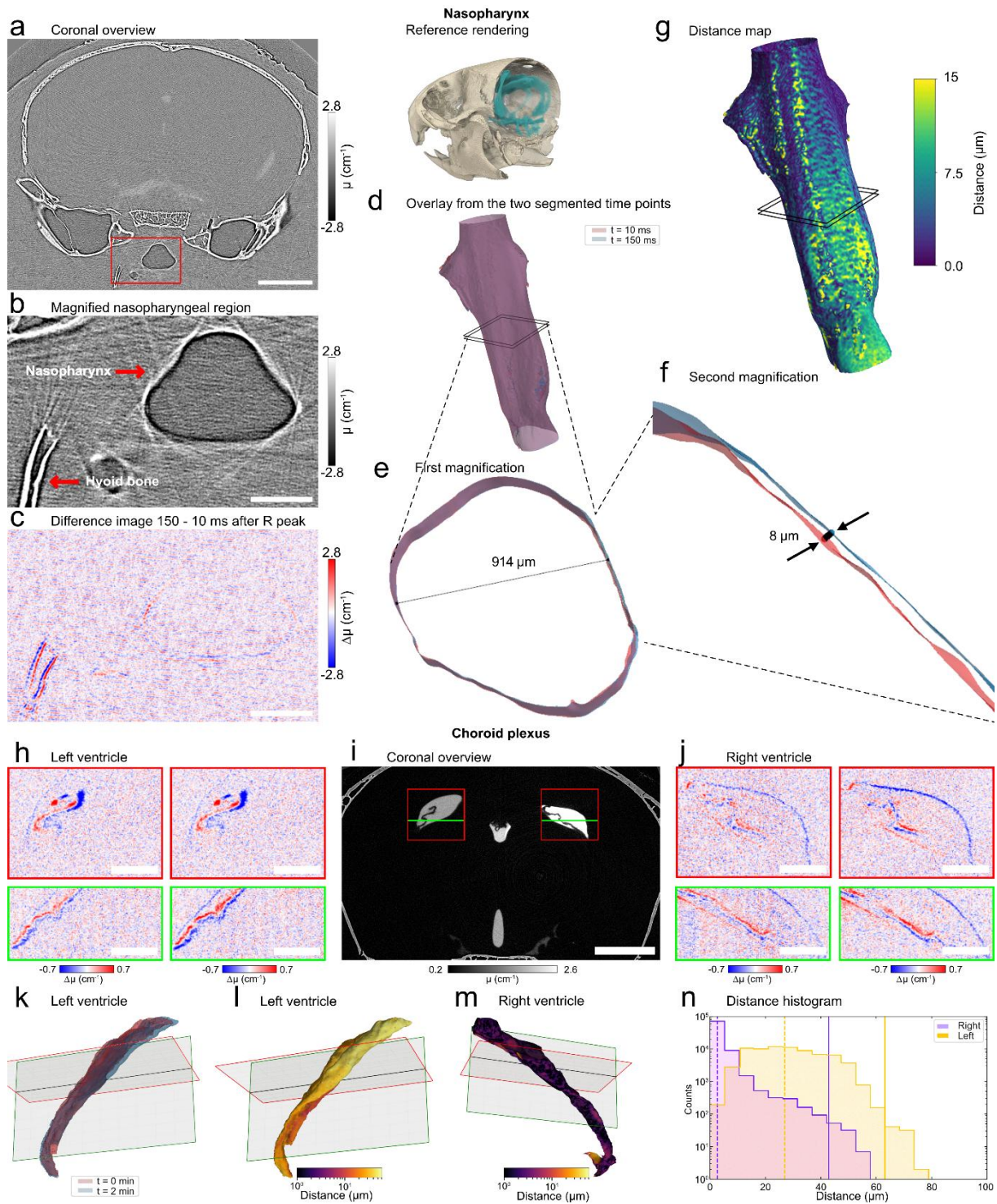

Quantification of intra- and extracranial tissue motion. (**a-g**) Nasopharynx: **a**, Coronal plane (left) - localized as shown in the rendering (right) - of a retrospectively cardiac-gated 3D acquisition, reconstructed with projections at 10 ms after the ECG R-peak. The frame region is magnified in **b**. µ: attenuation coefficient. Scale bar: 2 mm. **b**, Magnified view of nasopharynx and surrounding bone from **a**. **c**, Subtraction image of reconstructions performed with projections at 10 and 150 ms after R-peak. Red values indicate an increase in attenuation coefficient from 10 to 150 ms, blue a decrease. Scale bar: 0.5 mm. **e**, **f**, Manually segmented and smoothed surfaces of the nasopharynx at 10 (red) and 150 ms (blue) after R-peak. **e**, **f**, Magnified contour from the coronal cutting planes in **d**, with further magnified segment. **g**, Distance map between the nasopharyngeal surfaces in **d** at 10 and 150 ms after R-peak. (**h-n**) Choroid plexus (ChP): **h**, Subtraction images to inspect ChP movement between two acquisition time points in the left framed region (top row) and green section (bottom) in **i**. Difference between time points: 1 min (left column) and 2 min (right). Red values indicate an increase in attenuation coefficient in time, blue a decrease. Scale bars: 0.5 mm. **i**, Coronal plane with left lateral ventricle (left red frame) and right lateral ventricle (right red frame) marked for analysis in **h**, **j**. Green lines indicate horizontal section planes also used in those panels. Scale bar: 2 mm. **j**, Subtraction image as in **h**, but for the right lateral ventricle. **k**, Rendering of segmented left lateral ChP, acquisitions at two separate time points, 2 min apart. The red- and green-bordered planes correspond to those in **h-j**. **l**, Distance map between ChP at two time points as shown in **k**. Note the log-scaling of the colors. **m**, ChP distance map as in **l**, but for the right lateral ventricle. **n**, Histogram of distances shown in **l**, **m**, 21 bins in a range of [0 100] μm. Purple and yellow curves represent right and left lateral ventricular ChP, respectively. The dotted vertical lines indicate, for the respective lateral ventricle, median values (right: 3 μm, left: 27 μm), while the solid vertical lines mark the 99.9 percentile limits (right: 43 μm, left: 63 μm). Note the log-scaling of the vertical axis.

# Supplementary information

## Selection of contrast agent

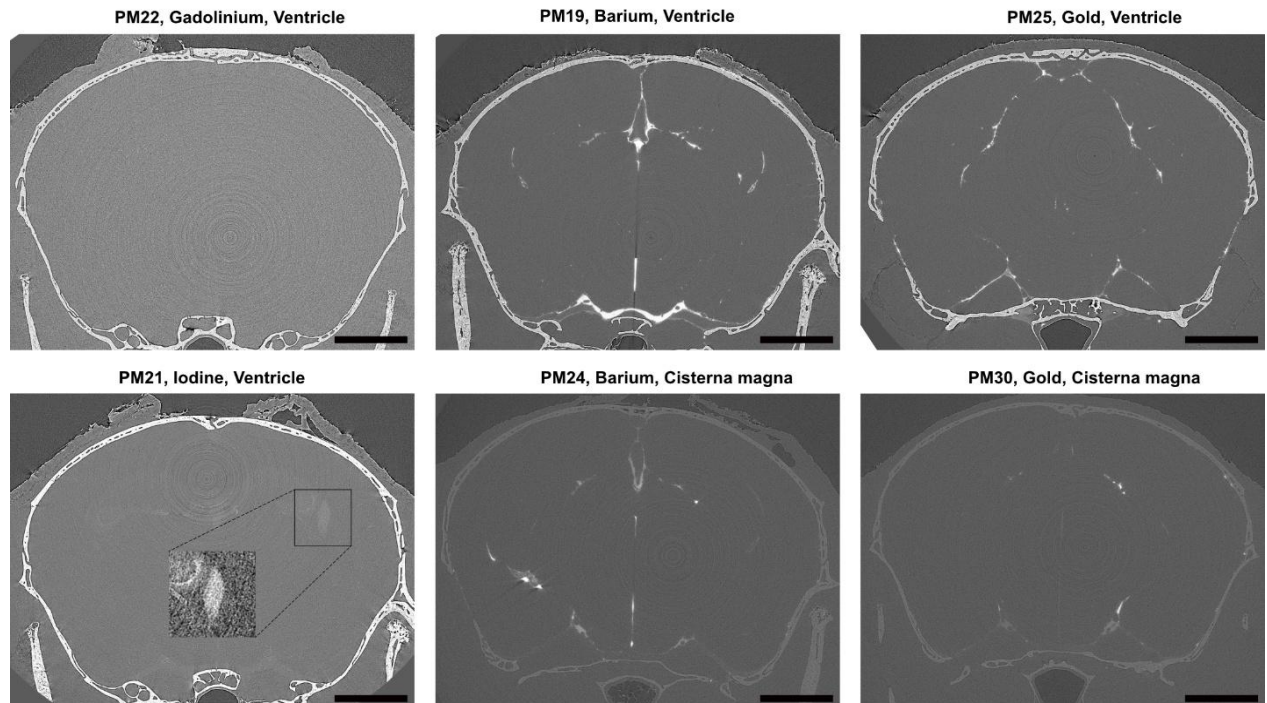

**Supplementary Figure 1.** Representative post-mortem images for contrast agent screening. Images illustrate the range of outcomes in contrast distribution following either ventricular or cisterna magna injections. Each SubjectID is given above the corresponding image. Attenuation coefficient values in the range of [-0.38 0.68] cm$^{-1}$ (for PM22 and PM21) or [-0.90 2.00] cm$^{-1}$ (for PM19, PM24, PM25 and PM30) were mapped onto an 8-bit dynamic range for better visualization in this figure. Scale bars: 2 mm.

# Fourier shell correlation

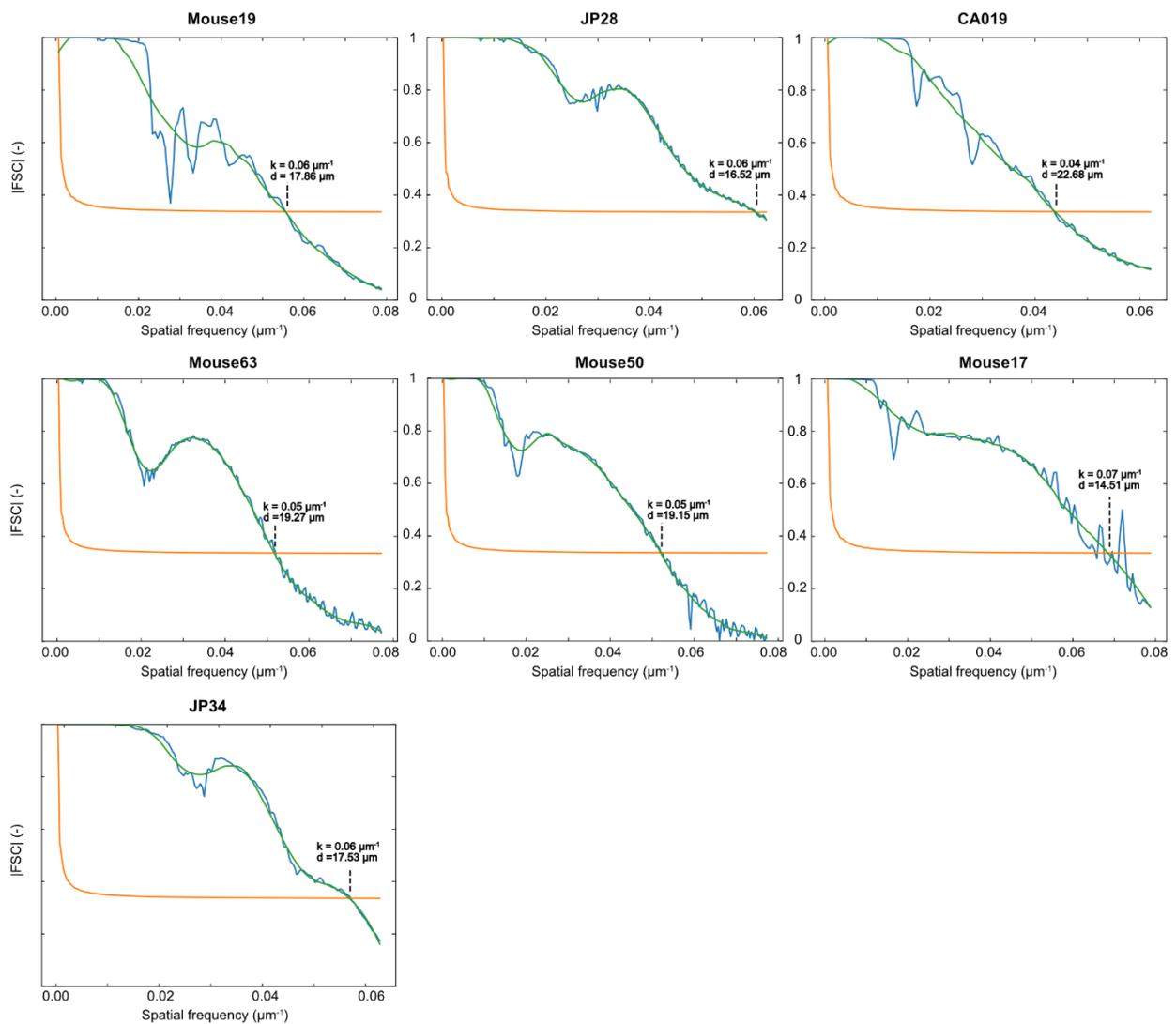

**Supplementary Figure 2.** Effective spatial resolution of the datasets in the manuscript. Fourier shell correlation (FSC) curves (green) and 1-bit threshold curves (orange) used to estimate spatial resolution as limited by signal-to-noise for datasets of Mouse19, Mouse63, JP28, Mouse50, CA019, Mouse17 and JP34, yielding resolutions of d = 17.9, 19.3, 16.5, 19.2, 22.7,14.5 and 17.5 μm, respectively.

**SUPPLEMENTARY TABLE 1.** EXPERIMENTAL AND IMAGING PARAMETERS

| SubjectID | Beamline | Infusion rate (μL/min) | Infused volume (μL) | Anesthesia regime[a] | Voxel size[b] (μm) | FOV[c] (mm) | Projections /scan (#) | Projection exposure time (ms) | Total scan acquisition time[d] (s) | Scan type |
|---|---|---|---|---|---|---|---|---|---|---|
| Mouse19 | ID17 | - | - | Ket/med | 6.3 | 16.1 x 4.9 | 2'000 | 5 | 20 | Single |
| JP28 | BL20B2 | 0.25 | 5 | Isoflurane | 7.99 | 16.4 x 12 | 1'800 | 5 | 23 | Single |
| CA019 | BMIT | 0.25 | 5 | Med/mid/but | 8 | 16.4 x 9.6 | 2'000 | 15 | 32 | Single[e], Time series |
| Mouse63 | ID17 | 0.5 | 2.5 | Ket/med | 6.45 | 16.5 x 4.5 | 2'000 | 4 | 10 | Time series |
| Mouse50 | ID17 | 0.2 | 1 | Ket/med | 6.45 | 16.5 x 4.5 | 2'000 | 4 | 10 | Time series |
| Mouse17 | ID17 | 0.5 | 5 | Ket/med | 6.3 | 16.1 x 4.9 | 60'000 | 5 | 600 | Gated |
| JP34 | BL20B2 | - | - | Med/mid/but | 7.92 | 16.2 x 11.9 | 1'800 | 4 | 22 | Time series |

[a]Anesthesia regime used during imaging. Analgesia and anesthesia induction regime are specified in the methods section and the FABRIC4 Portal. [b]Effective voxel size resulting from beamline-specific magnification optics and camera pixel size of 6.5 μm. [c]Field of view (FOV): horizontal x vertical. [d]Total scan acquisition time refers to total exposure and overhead time per scan, but excludes time needed to acquire reference images and to return the rotation stage to its original position. [e]A single scan was taken for the live state of peri-mortem imaging.

| SubjectID | Contrast agent | Compound type | Radiopaque element | K-edge energy (keV) | Injection site | Visual assessment[a] |
|---|---|---|---|---|---|---|
| PM20 PM26 | Iohexol | Low molecular weight organic molecule | Iodine | 33.2 | Ventricle Cisterna magna | No contrast |
| PM21 PM27 | Exitron P | High molecular weight organic molecule | Iodine | 33.2 | Ventricle Cisterna magna | Weak contrast |
| PM22 PM28 | Gadodiamide | Low molecular weight chelate complex | Gadolinium | 50.2 | Ventricle Cisterna magna | No contrast |
| PM23 PM29 | GadoSpin P | High molecular weight chelate complex | Gadolinium | 50.2 | Ventricle Cisterna magna | No contrast |
| PM19 PM25 | Exitron nano 12000 | Nanoparticles with hydrophilic coating | Barium | 37.4 | Ventricle Cisterna magna | Good contrast |
| PM24 PM30 | AuroVist 15 nm | Nanoparticles with hydrophilic coating | Gold | 80.7 | Ventricle Cisterna magna | Good contrast |

[a]Specimens were classified by visual assessment into three categories: no contrast (no visible intracranial contrast agent), weak contrast (contrast agent visible but contrast insufficient for *in vivo* imaging), and good contrast (sufficient contrast agent distribution and visibility for *in vivo* use).

## Data availability

Reconstructed 3D SRμCT datasets are available under a Creative Commons Attribution 4.0 International license via individual dataset entries on Zenodo, linked through the main Zenodo repository. For time series experiments, a single reconstructed 2D section is provided for each time point. Full 3D stacks are included for only one or two representative time points due to their cumulative size of 2.7 terabytes. The remaining datasets are available upon request. Computer-aided design (CAD) files of the mouse holder, infusion cannula, intra-cisterna magna infusion and tracheotomy stage, along with extended metadata and quantitative results tables, are also available in the main Zenodo repository. Information on other available datasets can be accessed via the FABRIC4 portal.

## Code availability

Code is available on Zenodo under GNU General Public License v3.0 or later. The repository includes Python scripts for calculating Fourier shell correlation (FSC) and Fourier ring correlation (FRC), MATLAB and elastix code for image registration, Python code using scikit-image along with Amira project files for image segmentation, and Python code using PyVista to generate meshes from the segmentations. It also includes MATLAB code for retrospective cardiac gating, LabVIEW code for animal monitoring equipment, Arduino code for ventilator synchronization, as well as Nabu scripts and Python code using TomoPy for tomographic reconstruction.

## Overview of animals and experiments

Supplementary Table 1 summarizes imaging parameters and mice reported on in the manuscript, including animal identifier (SubjectID), beamline, contrast agent infusion rate (μL/min), infused contrast agent volume (μL), anesthesia regime, voxel size (μm), field of view (mm), projections per scan (#), projection exposure time (ms), total acquisition time (s) and scan type (single, time series, or gated). For additional information on each mouse, please refer to the FABRIC4 portal.

# Selection of contrast agent

Six different types of commercially available X-ray contrast agents for preclinical angiography were evaluated for their performance in imaging cerebrospinal fluid spaces in the mouse brain. Contrast agents were injected *in vivo*, and the animals were imaged post-mortem at the Biomedical Beamline ID17 of the European Synchrotron Radiation Facility (ESRF).

## Methods

C57BL/6J mice (strain code 632) were supplied by Charles River Laboratoire France. To ensure proper acclimatization, the animals were housed at the beamline's animal facility for at least one week before the experiments. All procedures adhered to the European guidelines for animal experimentation (2010/63/EU). The experimental protocols were reviewed and approved by the responsible French ethics committee Comité d'éthique en expérimentation animale de l'ESRF (ETHAX), approval number APAFIS #30913-2021040211343677 v1.

12 female mice (age: 12 weeks; body weight range: 24-26 g) were first injected subcutaneously with buprenorphine (0.1 mg/kg) for analgesia. Anesthesia was induced 30 min after onset of analgesia via intraperitoneal injection of ketamine (73 mg/kg) and medetomidine (0.18 mg/kg). The anesthesia depth was monitored by testing reflexes, and additional injections of the same anesthetic mixture were given as needed. Eye ointment was applied, and the skull, neck, and upper thoracic region of the mouse were shaved to avoid imaging artifacts.

Before injection, the contrast agent to be tested (Supplementary Table 2) was carefully loaded into either a 34G/9.53 mm needle attached to a 10 μL Hamilton syringe (for intracerebroventricular injection) or into a cannula tube implant assembled with a Hamilton removable needle compression fitting (1 mm) attached to a 10 μL Hamilton syringe (for intra-cisterna magna injection). If bubbles were present in the syringe, the contrast agent was expelled completely and newly filled into the syringe.

### Intracerebroventricular injection of contrast agent

The animal was carefully positioned in a stereotactic frame on a warming pad. Blunt, non-perforating ear bars were used to position the animal, paying attention that the skull was aligned in a completely horizontal position. Core body temperature of the animal was monitored and maintained stable with aid of a rectal thermometer, thermal rescue blanket and heating pad. After the re-application of eye ointment and verification of sufficient anesthesia depth, a rostro-caudal skin incision was made to expose the skull. Excess periosteum from the bone was removed and bregma was identified. A small hole of about 1 mm diameter was drilled through the parietal bone at the injection coordinates, which were 0.95 mm lateral and 0.22 mm caudal of the bregma. The tip of the 34G Hamilton syringe needle was lowered into the hole to 2.35 mm depth, and 5 μl of contrast agent was injected at a rate of 1 μl/min over 5 min using a syringe pump. Following the injection, the needle was slowly removed after a 2 min waiting period, the skin was placed back over the injection site and sealed with tissue adhesive (e.g., Vetbond®).

## Cisterna magna (CM) injection

Monitoring and maintenance of the animal's core body temperature was done as described above. The animal was placed in prone position on a custom-made stereotactic platform and blunt, non-perforating ear bars were placed. Special care was taken to ensure that the skull was in a level position. The head of the mouse was then bent downwards at an angle of 90°.

A rostro-caudal incision was made into the skin at the back of the neck. The trapezius muscle was split along the midline. The long back and short neck muscle groups were then locally disconnected and bluntly prepared until the greyish atlanto-occipital membrane became visible. This membrane was pierced with a 26G needle and a 30G cannula-tube implant was quickly inserted into the cisterna magna and sealed with a drop of cyanoacrylate glue. The implant was connected to the infusion pump. A maximum of 10 µl cyanoacrylate accelerator was applied to the bonding surfaces to reduce the curing time. This required a waiting period of about 2 min. The head of the mouse was repositioned to level position. The skin was placed back over the neck and closed with tissue adhesive (e.g., Vetbond®). 5 µl of contrast agent was injected at a rate of 1 µl/min over 5 min.

After 30 min, the mice were euthanized using an overdose of anesthesia (ketamine (160 mg/kg) and medetomidine (0.4 mg/kg)), then fixed and stored in 4 % formaldehyde for 3 days. Before imaging, the bodies were briefly dabbed with laboratory tissue paper to remove excess liquid and mounted into a customized animal holder before being transferred to the experimental hutch of the ID17 beamline of ESRF, where they were imaged using a monochromatic beam at photon energies slightly above the K-edges of the respective contrast agents listed in Supplementary Table 2. Before reaching the monochromator, the polychromatic beam passed through a fixed filter stack of 0.8 mm thick graphite, one 0.5 mm aluminum sheet, and a second aluminum sheet of 2.0 mm thickness. A total of 4000 radiographs over a rotation range of 360° were acquired with a pco.edge 5.5 camera, coupled to serial Hasselblad 100 mm f/2.2 and Hasselblad 210 mm f/4 lenses, and a 250 µm LuAG:Ce scintillator, resulting in an effective pixel size of 3.1 µm[1]. The field of view was limited to 2560 × 2019 (7.9 mm × 6.3 mm) by the vertical extent of the beam. Radiographs were recorded with 100 ms exposure time and 1 ms overhead time. Acquisition time per scan was 6 min 44 s. The specimen-to-detector distance was 0.3 m.

Radiographs were reconstructed with a filtered back-projection algorithm implemented in PyHST2 software[2]. For mouse PM21, a 2D median filter with radius 5 pixels was employed to obtain the inset in Supplementary Figure 1.

## Results

For a contrast agent to be classified as suitable for *in vivo* experiments, it had to demonstrate sufficient distribution within the CSF space and provide adequate contrast to clearly visualize relevant anatomical structures, such as ventricles or subarachnoid space. Imaging post-mortem provided substantially more favorable conditions than imaging *in vivo*, as the specimens were static, eliminating motion artifacts that could have impacted the evaluation of contrast or recognition of smaller anatomic features. It also allowed for longer exposure times for higher signal-to-noise ratios.

Based on visual assessment, the specimens containing small molecular weight iodine or gadolinium-based contrast agents (0.8 kDa iohexol and 0.6 kDa gadodiamide) provided no visible intracranial contrast. PM22 is shown as a representative example of this category in Supplementary Figure 1. Small molecular weight contrast agents are known to rapidly extravasate when used in angiography. The lack of contrast may be explained by diffusion of the contrast agent out of the CSF spaces into surrounding tissue after injection.

In specimen PM21 injected with high molecular weight iodine-based ExiTron P contrast agent, structures such as the ventricles were difficult to discern and the specimen was classified as having weak contrast (Supplementary Table 2). This may be explained by more limited diffusion of the 20 kDa contrast agent compared to the 0.6 or 0.8 kDa small molecular weight contrast agents. The level of contrast was considered insufficient for the envisioned *in vivo* experiments, where much shorter exposure times well below the cardiac cycle length of mice would be employed. While it is likely that the concentration of the contrast agent dropped post-mortem during the 3 days of immersion in formaldehyde – meaning that the contrast agent could have had higher contrast if imaged shortly after injection in an *in vivo* imaging experiment – confirming this finding would have required another dedicated beamtime experiment. We also note that the ExiTron P we purchased and tested in 2020 was a different compound than the one currently sold under same name (as of July 2025). Instead of a 20 kDa linear polymer, the current version is a two orders of magnitude larger polymeric capsule with 290 nm hydrodynamic diameter, which we expect to yield results more in line with the nanoparticle-based contrast agents.

The high molecular weight gadolinium chelate complex-based GadoSpin P did not show visible contrast (data not shown). Due to its size of 200 kDa, diffusion out of the CSF spaces would not explain the lower contrast compared to ExiTron P. The lack of visible contrast may instead be explained by the lower concentration of 4 mg Gd/ml of the radiopaque compound in GadoSpin P compared to the concentration of 120 mg I/ml in ExiTron P.

Both gold and barium nanoparticle-based contrast agents were classified as having good contrast. The barium-based contrast agent ExiTron nano 12000 with 110 nm diameter nanoparticles (PM19, PM25) showed the most even distribution and highest signal intensity throughout the CSF space. The gold-based contrast agent AuroVist with 15 nm diameter particles (PM24, PM30) also provided good contrast, but presented with more agglomerates. This was observed in both ventricular (PM19 against PM24) and cisterna magna (PM25 against PM30) injections (Supplementary Figure 1). Consequently, the barium-based ExiTron nano 12000 was chosen for our *in vivo* imaging experiments.

# Anatomic imaging

## Supplementary Video 1

Peri-mortem shrinkage of native-state ventricular spaces in a mouse euthanized with an overdose of anesthetics (SubjectID: JP34). The movie shows a time-lapse of 60 reconstructed and registered coronal slices depicting the lateral ventricular region where contrast agent would be infused for contrast-enhanced imaging. Time between frames is 40 s. The video is rendered at 4 frames per second, corresponding to 160× speed. Recording begins immediately after anesthetic overdose. From 14 to 19 min (time points 20 to 29), movement artifacts are visible across the entire image, caused by body movement during unconscious deep final breaths before respiratory arrest. Following this period, progressive shrinkage of the ventricles can be observed, accompanied by a more pronounced choroid plexus movement compared to the live period at the beginning of the movie. Scale bar: 1 mm.

# Spatiotemporal solute distribution

## Supplementary Video 2

Mapping spatiotemporal solute distribution throughout the cranial CSF space following intracerebroventricular infusion – coronal reconstructions (SubjectID: Mouse63). The movie shows reconstructed and registered coronal slices of a plane intersecting the right lateral ventricle infusion site, see Fig. 4d (left). The time series consisted of 100 scans acquired every 30 s and shown as timelapse at 120× speed. By the 19[th] time point (9 min), the peak of attenuation coefficient in the right lateral ventricle was recorded as 7.5 cm$^{-1}$, see Fig. 4c (left). At this time, the contrast agent had filled the lateral ventricles and the third ventricle. Subsequently, contrast agent concentration in the right ventricle decreased until the end of the time series, with final measured values at the 100[th] time point of 1.5, 1.0 and 1.0 cm$^{-1}$ for the right lateral ventricle, the third ventricle and the left lateral ventricle, respectively. The dynamic profile depicted in this video is quantitatively represented in Fig. 4c (left), which shows the temporal evolution of the attenuation coefficients for selected regions of interest. Scale bar: 1 mm.

## Supplementary Video 3

Mapping spatiotemporal solute distribution throughout the cranial CSF space following intracerebroventricular infusion – projections (SubjectID: Mouse63). The movie shows non-registered radiographs from the starting position of the rotation stage at 0°. Acquisitions were made in 30 s-intervals and are shown as timelapse at 120× speed. In this view, the infusion cannula delivering the contrast agent, as well as contrast agent distribution throughout the lateral ventricles can be observed. After the 20[th] time point (9.5 min), the contrast agent in the infusion cannula is slowly pushed back into the infusion syringe due to backpressure. To counter this, a maintenance infusion at very low infusion rates after actual contrast agent delivery was introduced in later experiments. Scale bar: 1 mm.

## Supplementary Video 4

Mapping spatiotemporal solute distribution throughout the cranial CSF space following intra-cisterna magna infusion (SubjectID: Mouse50). The movie displays reconstructed and registered coronal slices of a plane intersecting the tip of the infusion cannula. The time series consisted of 50 scans each acquired every 30 s, shown at 120× speed. By the 7[th] time point (3 min), the maximum attenuation coefficient value in the cisterna magna was recorded as 4.5 cm$^{-1}$, see Fig. 4c (right). At this time, the contrast agent was also entering the subarachnoid space on either side of the cisterna magna. Subsequently, contrast agent concentration decreased to below 1.0 cm$^{-1}$ within 3 min. The dynamics depicted in this video are quantitatively represented in Fig. 4c (right), which shows the temporal evolution of the mean attenuation coefficient for selected regions of interest. Scale bar: 1 mm.

# Motion detection

## Supplementary Video 5

Quantifying extracranial tissue motion (SubjectID: Mouse17). Movie showing movement of the nasopharynx in coronal slices reconstructed with retrospective cardiac gating. Each frame represents a different phase in the cardiac cycle, with 10 ms interval between frames, shown at 1/25× speed. The most prominent movement can be observed during the first half of the movie. Scale bar: 1 mm.

## Supplementary Video 6

Quantifying intracranial tissue motion (SubjectID: CA019). Movie of orthogonal slices showing the left lateral ventricle after contrast agent arrival (13 min) from the infused right ventricle. Slow, non-periodic movement of the choroid plexus can be observed. Images were acquired every 1 min and are shown as timelapse at 240× speed, except between minutes 30 and 32 (time points 31 and 33), which are 2 min apart. Scale bar: 1 mm.